\documentclass[12pt]{article}
\usepackage{amsmath}
\usepackage{graphicx,psfrag,epsf}
\usepackage{enumerate}
\usepackage{natbib}
\usepackage{url} 
\usepackage{newtxmath}  
\usepackage{booktabs}
\newcommand{\blind}{0}

\begin{document}

\def\spacingset#1{\renewcommand{\baselinestretch}%
{#1}\small\normalsize} \spacingset{1}


\if0\blind
{
  \title{\bf Multiple Subordinated Modeling of Asset Returns}
  \author{Abootaleb Shirvani, Svetlozar T. Rachev, \\  Department of Mathematics and Statistics,Texas Tech University
  	\hspace{.2cm}\\
  	and \\
  	Frank J. Fabozzi \\
  	EDHEC Business School, United States}
  
  \maketitle
} \fi

\if1\blind
{
  \bigskip
  
 \begin{center}
    {\LARGE\bf Title}
\end{center}
  \medskip
} \fi

\bigskip
\begin{abstract}
Subordination is an often used stochastic process in modeling asset prices. Subordinated L\`{e}vy price processes and local volatility price processes are now the main tools in modern dynamic asset pricing theory. In this paper, we introduce the theory of multiple internally embedded financial time-clocks motivated by behavioral finance. To be consistent with dynamic asset pricing theory and option pricing, as suggested by behavioral finance, the investors’ view is considered by introducing an intrinsic time process which we refer to as a behavioral subordinator. The process is subordinated to the Brownian motion process in the well-known log-normal model, resulting in a new log-price process. The number of embedded subordinations results in a new parameter that must be estimated and this parameter is as important as the mean and variance of asset returns. We describe new distributions, demonstrating how they can be applied to modeling the tail behavior of stock market returns. We apply the proposed models to modeling S\&P 500 returns, treating the CBOE Volatility Index as intrinsic time change and the CBOE Volatility-of-Volatility Index as the volatility subordinator. We find that these volatility indexes are not proper time-change subordinators in modeling the returns of the S\&P 500.
\end{abstract}

\noindent%
\vfill

\newpage
\spacingset{1.45} 
\section{Introduction}
\label{sec:intro}

\noindent
There is a vast literature that has sought to model the dynamics of asset returns. The assumption typically made is that asset returns follow a normal distribution despite the preponderance of empirical evidence that rejects this distribution \citep[see][]{Rachev:2005}. There are several stylized facts about asset returns that should be recognized in modeling the dynamics of asset returns \citep[see][]{Cont:2001}.  Specifically, asset returns exhibit asymmetry and heavy tails. Modeling and analyzing the tail properties of asset returns are crucial for asset managers and risk managers. Consequently, the usefulness of the results of models that assume asset returns follow the normal law are questionable.

To deal with non-normality, the method of subordination \footnote{See \cite {Bochner:1955}, \cite{Sato:2002}, and \cite{Schoutens:2003}.} has been proposed in the literature to include business time and allow the variance of the normal distribution to change over time.
The subordination process in finance, also called random time change, $Y(t)=X\left(T(t) \right)$, under the assumption of independence of $X(t)$ and $T(t)$, is a technique employed to introduce additional parameters to the return model to reflect the heavy tail phenomena present in most asset returns and to generalize the classical asset pricing model. The concept of random time change was first applied to Brownian motion to obtain more realistic speculative prices by  \cite{Clark:1973}.  
\cite{Hurst:1997} applied various subordinated log-return processes to model the
leptokurtic characteristics of stock-index returns. They compared the classical log-normal model, the Mandelbrot and Fama log-stable model, the Clark model, the log symmetric generalized
hyperbolic model, the Barndorff--Nielsen model, the log hyperbolic model and the log variance gamma model in order to find the best three-parameter model that adequately takes into account leptokurtic characteristics for indices.\footnote{For a further discussion of the use of subordinators in financial modeling, see the books by \cite{Sato:2002} and \cite{Schoutens:2003}.} 
In the option pricing literature, \cite{Carr:2004} used random-time change to derive a more realistic price process and to extend the approach in \cite{Carr:2003} by providing an efficient way to include the correlation
between the stock price process and random-time change.
\cite{Klingler:2013} introduced two new six-parameter processes based on time changing
tempered stable distributions and developed an option pricing model based on these processes.

According to behavioral finance theory, the views of investors change the underlying asset process models. Investors view positive and negative returns on financial assets differently according to the disposition effect (i.e., the manner in which investors treat capital gains). Thus to obtain more realistic asset prices, it is essential to incorporate the views of investors in log-return and option pricing models.  To be consistent with dynamic asset pricing theory, the views of investors can be taken into account by introducing an intrinsic time process, what we refer to as a behavioral subordinator.  The process is subordinated to the Brownian motion process in the well-known log-normal model, resulting in a new log-price process.

In this paper, we define multiple subordinated methods, provide a model for the dynamics of asset returns, and generalize the classical log-normal asset pricing model. We do so by replacing physical time in the well-known return model by multiple stochastic intrinsic times, which allows for tail effects. In a double subordinated model, we view $\left\lbrace  V(t)=T(U(t)), t\geq0\right\rbrace$  as stock intrinsic time, and $\left\lbrace  U(t), t\geq0\right\rbrace  $  as stock-volatility intrinsic time or volatility subordinator. 

We will define and investigate the properties of various multiple subordinated log-return processes that are applied to model the leptokurtic characteristics of asset returns. The possible multiple subordinated models that we consider for the
distribution of changes in asset returns are the  $\alpha$-stable, gamma and inverse Gaussian subordinated models. 
These models differ by their intrinsic time processes that are subordinated to the standard Brownian motion for modeling asset returns. They are simple models
that usually add some extra parameters and reflect several fundamental probabilistic
relationships such as asymptotic laws, self-similarity, and infinite divisibility. There are other classes of distributional models that could be used, but they are more complex and do not emphasize characteristics that arise from fundamental relationships. Also, we generalize the multiple $\alpha$-stable subordination to the $\tau$-subordination or continuous subordination, arguing that $\tau$ could be an additional parameter that allows for heavy-tailedness in the modeling of asset returns. We show that two popular stock market volatility indexes -- the CBOE volatility index (VIX \footnote{ VIX is an index created by the CBOE, representing 30-day implied volatility calculated by S\&P 500 options. (see\ http://www.cboe.com/vix).}) and the CBOE volatility of volatility index (VVIX \footnote{The VVIX is an index created by the CBOE, (see http://www.cboe.com/products/vix-index-volatility/volatility-on-stock-indexes/the-cboe-vvix-index/vvix-whitepaper). It is a volatility of volatility (vol-of-vol) measure, and represents 30-day implied volatility calculated from  VIX options.}) -- are not proper intrinsic time change subordinators for modeling the stock market as measured by the SPDR S\&P 500 \footnote{See SPDR S\&P 500 ETF Indices, https://us.sprdrs.com/.} (an exchange-traded fund).

The remainder of this paper is organized as follows. In Section 2, we introduce the double subordinated model and present a multiple subordinated model using a continuous-time change process. In Section 3, we empirically estimate the return distribution of the stock market index by applying the double subordinated models that we presented in Section 2 and offer some concluding remarks in Section 4.

\section{Doubly Subordinated Price processes}
\noindent
Consider a stock price process $\left\lbrace S_t,\, t\ge0,\, S_0>0\right\rbrace ,$ with dynamics given by its log-price process $L_t=lnS_t,$

\begin{equation}
\label{Doubly_subordinated}
L_t=L_0+\mu t+\,\gamma U\left(t\right)+\,\rho T\left(U\left(t\right)\right)+\sigma B_{T(U\left(t\right))},\,t\ge0, \mu\in R,\gamma \in R,\rho\in R,\sigma >0, 
\end{equation} 
where the triplet $\left(B_s,T\left(s\right),\ U\left(s\right),s\ge 0\right)$ are independent processes generating stochastic basis $\left(\mathrm{\Omega },\mathcal{F},\ \mathbb{F}=\left({\mathcal{F}}_t,t\ge 0\right), \mathbb{P}\mathrm{\ }\right)$ representing the natural world with $\{   B_s, s\ge0 \} $ being a standard Brownian motion. $\{ T\left(s\right),\ U\left(s\right),s\ge0 \} $ and  $\{ T\left(0\right)=0,\ U\left(0\right)=0\} $  are L\'{e}vy subordinators. A L\textrm{\'{e}}vy subordinator is a L\textrm{\'{e}}vy process with increasing sample path.\footnote{See Chapter 6 in \cite{Sato:2002}.} 

$B_t$, $T\left(t\right)$ and $U\left(t\right)$ are ${\mathcal{F}}_t$-adopted processes whose trajectories are right-continuous with left limits. We view $V\left(t\right)=T\left(U\left(t\right)\right)$, $t\ge0$  as \textit{stock intrinsic time}, and $U\left(t\right)$, $t\ge 0$ as the \textit{stock-volatility intrinsic time} or the \textit{volatility subordinator}. For example, in modeling the SPDR S\&P 500 by the triplet  $\left(L_t,V\left(t\right),U\left(t\right)\right)$, $t\ge 0$ one can choose: $\left(i\right)\ L_t$, $t\ge0$ as a stochastic model for the SPDR S\&P 500  index;  $\left(ii\right)\ V\left(t\right)$, $t\ge0$ as the cumulative VIX (i.e., $V\left(t\right)$ representing the cumulative value of VIX in $\left[0,t\right]$) and $\left(iii\right)\,U\left(t\right)$ $t\ge0$ as the cumulative VVIX (i.e., $U\left(t\right)$, $t\ge0$ representing the cumulative value of VVIX in $\left[0,t\right]$).

The general framework of behavioral finance provides an alternative view of the doubly subordinated price process.\footnote{See Barberis and Thaler (2005).} In their seminal paper, \cite{Tversky:1992} introduced the Cumulative Prospect Theory (CPT). According to this theory, positive and negative returns on financial assets are viewed differently due to the general \textquotedblleft fear\textquotedblright  disposition of investors. To quantify an investor's fear disposition, \cite{Tversky:1992} and \cite{Prelec:1998} introduced a probability weighting function (PWF), $w^{\left(\mathcal{R},\mathcal{S}\right)}:\left[0,1\right]\to \left[0,1\right]$, transforming the asset return distribution $F_{\mathcal{R}}\left(x\right)\mathrm{=}\mathbb{P}\left(\mathcal{R}\le x\right),x\in R$ according to the investor's views to a new one  $F_{\mathcal{S}}\left(x\right)\mathrm{=}\mathbb{P}\left(\mathcal{S}\le x\right)=w^{\left(\mathcal{R},\mathcal{S}\right)}\left(F_R\left(x\right)\right),x\in R.$ Tversky and Kahneman (1992) introduced the following PWF

\begin{equation}
\label{PWT}
w^{\left(\mathcal{R},\mathcal{S};TK\right)}\left(u\right)=\frac{u^{\gamma }}{{\left[u^{\gamma }+{\left(1-u\right)}^{\gamma }\right]}^{\frac{1}{\gamma }}},\,\,u\in \left(0,1\right),\,\, \gamma \in \left[0,1\right]. 
\end{equation} 

Unfortunately, this choice of the PWF is inconsistent with dynamic asset pricing theory (DAPT) because $F_{\mathcal{S}}$ is not an infinitely divisible distribution function, leading to arbitrage opportunities in behavioral asset pricing models. \cite{Prelec:1998} introduced an alternative WPF:
\begin{equation}
\label{WPF}
w^{\left(\mathcal{R},\mathcal{S};P\right)}\left(u\right)={{\mathrm{exp} \left(-\delta lnu\right)\ }}^{\rho },\,\,u\in \left[0,1\right],\,\,\delta >0,\,\,\rho \in \left(0,1\right).
\end{equation} 

\noindent Prelec's $w^{\left(\mathcal{R},\mathcal{S};P\right)}$ is consistent with DAPT only in the case when $\mathcal{R}$ has a Gumbel distribution given by 
\begin{equation}
\label{Gumbel}
F_{\mathcal{R}}\left(x\right)={\mathrm{exp} \left(-e^{-\frac{x-\mu }{\beta }}\right)\ },\,\,x\in R,\,\,\mu \in R,\,\,\beta >0. 
\end{equation}

\cite{Rachev:2017} studied the general form of PWF consistent with DAPT. Following their arguments, we view the $\mathcal{R}$ as the return in unit time of a single subordinated log-price process; that is,  $\mathcal{R}=M_1,$ where
\begin{equation}
\label{single_subordinated}
M_t=lnS_0+\mu t+\gamma U\left(t\right)+\sigma B_{U(t)}\, ,\,\,t\ge 0,\,\,\mu \in R,\,\,\gamma \in R,\,\,\sigma >0. 
\end{equation}

The log-price process $M_t$, $t\ge0$ represents the asset price dynamics before the introduction of the views of investors. Investor's fear disposition amounts to the introduction of a second (\textquotedblleft behavioral\textquotedblright) subordinator $T\left(t\right)$, resulting in a new log-price process, 

\begin{equation}
\label{Double_logprice_process}
L_t=lnS_0+\mu t+\gamma U\left(t\right)+\rho T\left(U\left(t\right)\right)+\sigma B_{T(U\left(t\right))},\,\,t\ge0,\,\,\rho \in R.
\end{equation} 

The distribution of  $\mathcal{S}=L_1$ is characterized by heavier tails than $\mathcal{R},$ representing the general fear disposition of the investor. The corresponding WPF, $w^{\left(\mathcal{R},\mathcal{S}\right)}:\left[0,1\right]\to \left[0,1\right]$,  is defined by $w^{\left(\mathcal{R},\mathcal{S}\right)}\left(u\right)=F_{\mathcal{S}}\left(F^{inv}_{\mathcal{R}}\left(u\right)\right)$ where $F^{inv}_{\mathcal{R}}\left(u\right)={\mathrm{min} \left\{x:F_{\mathcal{R}}\left(x\right)>u\right\}\ }$ is the inverse function of $F_{\mathcal{R}}\left(x\right)$.\footnote{The corresponding WPF, $w^{\left(\mathcal{R},\mathcal{S}\right)}:\left[0,1\right]\to \left[0,1\right]$, is defined by   $w^{\left(\mathcal{R},\mathcal{S}\right)}\left(u\right)=F_{\mathcal{S}}\left(F^{inv}_{\mathcal{R}}\left(u\right)\right)\ $ where $F^{inv}_{\mathcal{R}}\left(u\right)={\mathrm{min} \left\{x:F_{\mathcal{R}}\left(x\right)>u\right\}}$ is the inverse function of $F_{\mathcal{R}}\left(x\right)$.  $F_{\mathcal{S}}\left(x\right)=w^{\left(\mathcal{R},\mathcal{S}\right)}\left(F_R\left(x\right)\right)$,  with  $F_R\left(x\right)=u,\ x=F^{inv}_R\left(u\right)$, and from,   $F_{\mathcal{S}}\left(x\right)=w^{\left(\mathcal{R},\mathcal{S}\right)}\left(F_R\left(x\right)\right)$, we have  $F_{\mathcal{S}}\left(F^{inv}_R\left(u\right)\right)=w^{\left(\mathcal{R},\mathcal{S}\right)}\left(u\right)$.}
In this setting, the log-price parameters for $M_t$, $t\ge0$ should be estimated from the spot prices of the underlying stock. We view $M_t$, $t\ge0$ as the dynamics of the log-price process $lnS_t$ as observed by spot traders at the current time, $t=0$. Thus, the parameters of $M_t$, $t\ge0$ are estimated from the spot market. However, we consider $L_t$, $t\ge0$ as the dynamics of the log-price process $lnS_t$ as seen by option traders. The motivation for this choice for the doubly subordinated process  $L_t$, $t\ge0$ is the generally accepted view that option traders are more \textquotedblleft fearful\textquotedblright than spot traders due to the non-linearity of the risk factors they face. Therefore, the remaining parameters, $\rho\in R$, and the parameters for the distribution of $T\left(1\right)$ and $U\left(1\right)$ should be calibrated from the risk-neutral dynamics $L^{risk-neutral}_t$, preserving the double subordinated structure of  $L_t$.\footnote{ See change of measure theorem for L\textrm{\'{e}}vy processes in Chapter 6 in \cite{Sato:2002} and Chapter 3 in \cite{Jacod:2003}.}

\subsection{Double-Stable subordinators and Normal-Double-Stable log-price processes} 

\noindent \cite{Mandelbrot:1967} were the first to apply a subordinated Brownian motion to modeling asset returns. In their model, the log-price process is modeled by
	\begin{equation}
	\label{subordinated_Brownian_motion}
	L_t=L_0+\mu t+\rho T\left(t\right)+\sigma B_{T(t)},\,\,t\ge0,\,\,\mu\in R,\,\,\rho\in R,\,\,\sigma>0,
	\end{equation} 
	where L\'{e}vy subordinator $T\left(t\right)$, $t\ge0$ is $\frac{{\alpha }_T}{2}$\textit{-stable subordinator} \citep[see][Proposition 1.3.1]{Samorodnitsky:1994}  for ${\alpha }_T\in \left(0,2\right)$ independent of the Brownian motion $B_t$, $t\ge0$. The unit increment of $T\left(t\right)$, $t\ge0$ has Laplace transform
	
	\begin{equation}
	\label{Laplace_transform_stable}
	\mathfrak{L}_{T\left(1\right)}\left(s\right)=\mathbb{E}e^{-sT\left(1\right)}={\mathrm{exp} \left(-{({\delta }_Ts)}^{\frac{{\alpha }_T}{2}}\right)\ },\ s>0,\ {\delta }_T>0,
	\end{equation} 
	where parameter ${\delta }_T>0$ is a scale parameter and $\frac{{\alpha }_T}{2}$ is the tail index \footnote{ See \cite{Samorodnitsky:1994} for more information about stable random variables and stable processes that we use in this paper.}. The tail-probability function $S_{T\left(1\right)}\left(x\right)\mathrm{=}\mathbb{P}\left(T\left(1\right)>x\right)$, $x\ge0$ is regularly varying (RV) \footnote{Recall that a function $f:(0,\infty)\to(0,\infty)$ is called \textit{regularly varying} (at infinity) of order $r\in R$, denoted $f\in RV\left(\alpha\right)$, if $f\left(x\right)=x^rL\left(x\right)$, where $L:(0,\infty)\to (0,\infty )$ is a slowly varying function (at infinity); that is,  ${\mathop{\mathrm{lim}}_{x\uparrow \infty} \frac{f\left(bx\right)}{f\left(x\right)}=1}$ for every $b>0$. For $f\in RV\left(r\right)$, we call $(-r)$ the tail index of $f$, and denote it by $f\in TI\left(-r\right)$.} of order $\left(-\frac{{\alpha }_T}{2}\right)>-1$ (see Samorodnitsky and Taqqu (1994), Proposition 1.2.15), and  thus, $S_{T\left(1\right)}\in TI\left(\frac{{\alpha }_T}{2}\right).$ The explicit form of all moments -- $\mathbb{E}\left(T{\left(1\right)}^p\right)<\infty ,\ 0<p<\frac{{\alpha }_T}{2}$ -- is given in \citet[p.18]{Samorodnitsky:1994}.  \cite{Tagliani:2004} provide a numerical procedure to approximate (in total variation distance) the density $f_{T\left(1\right)}\left(x\right)$, $x>0$ if there are a sufficient number of moments 
	\begin{equation}
	\label{moments_stable}
	\mathbb{E}\left(T{\left(1\right)}^{p_j}\right)=\frac{p_j}{\mathrm{\Gamma }(1-p_j)}\int^{\infty }_0{\frac{1-{\mathfrak{L}}_{T\left(1\right)}\left(s\right)}{s^{p_j+1}}}ds,\,\,\, 0<p_j<\frac{{\alpha }_T}{2},\,\,\ j=1,\dots ,J, 
	\end{equation} 
	are given. In other words, if the sample moments of order  $p_j\in \left(0,\frac{{\alpha }_T}{2}\right),\ j=1,\dots ,J$ are available and $J$ is sufficiently large, we can approximate the probability density function (pdf) $f_{T\left(1\right)}$ in ${\mathcal{L}}_1$-distance (total variation distance).
	
	The subordinated Brownian motion, denoted by $B_{T(t)}$, is a ${\alpha}_T$-stable motion with unit increment with $B_{T(1)}$ having characteristic function (Ch.f.) given by \footnote{The proof is provided in Appendix A.1 in the supplementary material.}
	
	\begin{equation}
	\label{chf_stable}
	{\varphi }_{B_{T\left(1\right)}}\left(u\right)=\mathbb{E}{\mathrm{exp} \left\{iuB_{T\left(1\right)}\right\}\ }={\mathrm{exp} \left(-{\left(\frac{{\delta }_T}{2}\right)}^{\frac{{\alpha }_T}{2}}u^{{\alpha }_T}\right).}
	\end{equation}  
	That is, $B_{T(t)}$, $t\ge0$ is ${\alpha}_T$- stable motion with scale parameter ${\left(\frac{{\delta}_T}{2}\right)}^{\frac{{\alpha }_T}{2}}$, and thus, $S_{|B_{T\left(t\right)}|}\left(x\right)=\mathbb{P}\left(|B_{T\left(t\right)}|>x\right)$, $x\ge0$ is $RV\left({\alpha }_T\right)$. Consider now the price process model \eqref{Doubly_subordinated} with two stable subordinators  $T\left(t\right)$, $t\ge0$ with  unit increment Laplace transformation given by \eqref{Laplace_transform_stable} and $U\left(t\right)$, $t\ge0$ with
	
	\begin{equation}
	\label{increment_distribution_double_stable}
	{\mathfrak{L}}_{U\left(1\right)}\left(s\right)=\mathbb{E}e^{-sU\left(1\right)}={\mathrm{exp} \left(-{({\delta }_Us)}^{\frac{{\alpha }_U}{2}}\right)}\, ,\,\,s>0\,,\,\, {\delta }_U>0.
	\end{equation}
	Then, the L\'{e}vy subordinator $V\left(t\right)=T\left(U\left(t\right)\right)$, $t\ge0$ has unit increment  $V\left(1\right)$ with Laplace transform
	\begin{equation}
	\label{pdf_double_stable}
	{\mathfrak{L}}_{V\left(1\right)}\left(s\right)=\int^{\infty }_0{{\mathrm{exp} \left(-u{\left({\delta }_Ts\right)}^{\frac{{\alpha }_T}{2}}\right)\ }}f_{U\left(1\right)}\left(u\right)du.
	\end{equation}
	
	The only one known explicit form for  $f_{U\left(1\right)}$ is when $U\left(t\right)$, $t\ge0$ is a L\'{e}vy stable subordinator with scale parameter $b_U>0$. The pdf of $U\left(1\right)$ is given by
	\begin{equation}
	\label{pdf_stable}
	f_{U\left(1\right)}\left(x\right)=\sqrt{\frac{b_U}{2\pi }}x^{-\frac{3}{2}}{\mathrm{exp} \left(-\frac{b_U}{2x}\right)\ },\ x>0.
	\end{equation}
	\noindent In this case, the Laplace transform of $V(1)$ has the following representation
	\begin{equation}
	\label{Laplace_double_stable}
	{\mathfrak{L}}_{V\left(1\right)}\left(s\right)={\mathrm{exp} \left(-\sqrt{2b_u}{\left({\delta }_Ts\right)}^{\frac{{\alpha }_T}{4}}\right)}.
	\end{equation}
	\noindent That is, $V\left(1\right)$ is $\frac{{\alpha }_T}{4}$-stable subordinator.\footnote{ We shall often have the probability distributions of $U\left(1\right)$, $T\left(1\right)$, $V\left(1\right)$, $B_{T\left(t\right)}$, $B_{V\left(t\right)}$, and  $L_1$ in closed form in terms of their characteristic functions, Laplace transforms, or moment-generating functions. We will not discuss particular estimation procedures.  The estimation procedures are well studied in the literature that deals with estimating distributional parameters, probability density function  and cumulative distributions. The probability density function is recovered by using characteristic functions, Laplace transforms, and moment-generating functions. See, for example, \cite{Abate:1999}, \cite{Glasserman:2010}, \cite{Tsionas:2012}, \cite{Mnatsakanov:2013}, \cite{Carrasco:2017}, and \cite{Kateregga:2017}.} Therefore, the subordinated Brownian motion $B_{V\left(t\right)}$, $t\ge0$ is a  $\frac{{\alpha }_T}{4}$-stable motion.
	We refer to the L\'{e}vy subordinator  $V\left(t\right)=T(U\left(t\right))$, $t\ge 0$ with Laplace transform ${\mathfrak{L}}_{V\left(1\right)}$  given by \eqref{Laplace_double_stable} as the \textit{double-stable subordinator}. We shall call   $L\left(t\right)$, $t\ge0$ in \eqref{Double_logprice_process} the \textit{normal-double-stable log-price process}. The subordinated Brownian motion $B_{V\left(t\right)}$, $t\ge0$ is, therefore, a  $\frac{{\alpha }_T}{4}$-stable motion.\\

	Now let's look at the distribution of the \textit{normal-compound-stable log-price process} $L_t=lnS_t$, $t\ge0$ 
	given by 
	\[L_t=L_0+\mu t+\gamma U\left(t\right)+\rho T\left(U\left(t\right)\right)+\sigma B_{T(U\left(t\right))},\,t\ge0,\,\mu\in R,\,\gamma \in R,\,\rho \in R,\,\sigma >0.\] 
	The triplet $\left(B_s, T\left(s\right), U\left(s\right), s\ge0\right)$, 
	$U\left(1\right)\sim \text{L\textrm{\'{e}}vy-stable}\left(b_U\right)$ , $T\left(1\right)\sim\text{L\textrm{\'{e}}vy-stable}\left(b_T\right)$ 
	are independent processes generating stochastic basis  $(\mathrm{\Omega},\mathcal{F},  \mathbb{F}=\left({\mathcal{F}}_t,t\ge0\right),\mathbb{P})$
	representing the natural world. $B_s$, $s\ge0$ is a standard Brownian motion, and $T\left(s\right)$, $U\left(s\right)$, $s\ge0$, $\left( T\left(0\right)=0, U\left(0\right)=0 \right)$  are L\'{e}vy subordinators. Denote 
	\begin{eqnarray}
	\label{log_return_eq}
	\mathrm{\Lambda }\mathrm{:}=L_1-L_0=\mu +\gamma U\left(1\right)+\rho V(1)+\sigma B_{V(1)}.
	\end{eqnarray}
	\noindent The pdf of $\mathrm{\Lambda }$ is given by
	\begin{equation}
	\label{dist_double_stable}
	f_{\mathrm{\Lambda }}\left(x\right)=\frac{2\rho \sqrt{b_Tb_U}}{\sigma {\left(2\pi \right)}^{\frac{3}{2}}}\int^{\infty }_0{\frac{e^{\frac{\left(x-\mu -\gamma u\right)\rho }{{\sigma }^2}-\frac{b_U}{2u}}{\mathrm{K}}_1\left(\frac{\rho }{{\sigma }^2}\sqrt{{\left(x-\mu -\gamma u\right)}^2+b_T{\sigma }^2u^2}\right)}{\sqrt{u}\sqrt{{\left(x-\mu -\gamma u\right)}^2+b_T{\sigma }^2u^2}}du,}
	\end{equation}
	
	\noindent where ${\mathrm{K}}_n\left(x\right)$ is the modified Bessel function of the second kind.  The Ch.f. ${\varphi }_{\mathrm{\Lambda }}\left(v\right)=\mathbb{E}e^{iv\mathrm{\Lambda }},v>0\ $is given by \footnote{The proof is provided in Appendix A.2 in the supplementary material.}
	\begin{equation}
	\label{chf_double_stable}
	{\varphi }_{\mathrm{\Lambda }}\left(v\right)=\mathbb{E}e^{iv\mathrm{\Lambda }}=e^{iv\mu }{\mathrm{exp} \left\{-\sqrt{-2b_U\left(iv\gamma -\sqrt{-2b_T\left(iv\rho -\frac{1}{2}v^2{\sigma }^2\right)}\right)}\right\} }.
	\end{equation}
	We call $\mathrm{\ }\mathrm{\Lambda }$-distribution the \textit{normal-compound-L\'{e}vy-stable distribution}, and   $L\left(t\right)$, $t\ge0$ a \textit{normal-compound-stable log-price process.} We note that the moments of $\mathrm{\Lambda }$ are undefined.
	
	Here is an example of multiple L\'{e}vy stable subordinations. Let $U^{\left(n\right)}\left(t\right)$, $t\ge0$ be L\'{e}vy stable subordinators with scale parameter $b_n>0$; that is, $U^{\left(n\right)}\left(1\right)\sim \text{L\textrm{\'{e}}vy-stable}\left(b_n\right)$. Set  $V^{\left(1\right)}\left(t\right)=U^{\left(1\right)}\left(t\right)$ and $V^{\left(n\right)}\left(t\right)=U^{\left(n\right)}\left(V^{\left(n-1\right)}\left(t\right)\right)$ for $n=2,3,\dots $ Then, the Laplace exponent of  $V^{(n)}\left(t\right)$ is given by \footnote{The proof is provided in Appendix A.3 in the supplementary material.} 
	\begin{equation}
	\label{laplaec_exponet_n_sub}
	{\mathrm{\Phi }}_{V^{\left(n\right)}}\left(s\right)=s^{2^{-n}}\prod^n_{k=1}{{\left(2b_k\right)}^{2^{-k}}},\,\,s>0,\ n\in \mathcal{N}=\left\{1,2,\dots .\right\}.
	\end{equation}
	
	Letting  $n\uparrow \infty $ \footnote{\noindent The distributional tail of $V^{(n)}\left(1\right)$ becomes heavier and heavier as $n\uparrow \infty$. In the limit, if ${\mathop{\mathrm{sup}}_{n\in \mathcal{N}} \prod^n_{k=1}{{\left(2b_k\right)}^{2^{-k}}}\ }<\infty$ then ${\mathop{\mathrm{lim}}_{n\uparrow \infty } {\mathfrak{L}}_{V^{(n)}\left(1\right)}\left(s\right)\ }={\mathop{\mathrm{lim}}_{n\uparrow \infty } {\mathrm{exp} \left(-s^{2^{-n}}\prod^n_{k=1}{{\left(2b_k\right)}^{2^{-k}}}\right)\ }\ }={\mathrm{exp} \left(-s^0\right)\ }={\mathrm{exp} \left(-1\right)}$.}  
	and assuming that  ${\mathop{\mathrm{sup}}_{n\in \mathcal{N}} \prod^n_{k=1}{{\left(2b_k\right)}^{2^{-k}}}\ }<\infty$, the distribution of  $V^{\left(n\right)}\left(1\right)$ degenerates as the distributional mass of $V^{\left(n\right)}\left(1\right)$ escapes  to infinity as $n\uparrow \infty$. As the tail-probability function $S_{V^{\left(n\right)}\left(1\right)}\in TI\left(2^{-n}\right)$, then the random variable ${\xi }^{\left(n,\beta \right)}=V^{\left(n\right)}{\left(1\right)}^{\frac{2^{-n}}{\beta }}$, $\beta>0$ will be: $\left(a\right)$ in the domain of attraction of $\beta$-stable random variable if $\beta<2$, and $\left(b\right)$ in the domain of attraction of a normal law if $\beta\ge2$.\footnote{\noindent Since the tail-probability function  $S_{V^{(n)}\left(1\right)}\left(x\right)$ is $RV\left(2^{-n}\right)$, then $ {\xi }^{\left(n,\beta \right)}=V^{\left(n\right)}{\left(1\right)}^{\frac{2^{-n}}{\beta }}$, $\beta>0$ will be in the domain of attraction of a $\beta$-stable random variable if $\beta<2$, and in the domain of attraction of a normal law if $\beta\ge2$.We see   $\mathbb{P}\left(V^{\left(n\right)}\left(1\right)>x\right)=x^{-2^{-n}}L\left(x\right)\mathrm{,}$ and
		
		\noindent$\mathbb{P}\left({\xi }^{\left(n,\beta \right)}=V^{\left(n\right)}{\left(1\right)}^{\frac{2^{-n}}{\beta }}>x\right)=\mathbb{P}\left({\left(V^{\left(n\right)}{\left(1\right)}^{2^{-n}{\beta }^{-1}}\right)}^{2^n}>x^{2^n}\right)=
		\mathbb{P}\left({\left(V^{\left(n\right)}\right)}^{{\beta }^{-1}}>x^{2^n}\right)\\=\mathbb{P}\left(V^{\left(n\right)}>x^{2^n\beta }\right)={\left(x^{2^n\beta }\right)}^{-2^{-n}}L\left(x^{2^n\beta }\right)=x^{-\beta }L(x)$.} 
	
	Next we define a continuous version of multiple L\'{e}vy stable subordinators.  Let $b_n=B$, $n\in \mathcal{N}=\left\{1,2,\dots .\right\}$. We now extend \eqref{laplaec_exponet_n_sub} as follows: for every $\tau\ge\ 0$, define $V^{\left(\tau\right)}\left(t\right)$, $t\ge0$ as the L\'{e}vy process with Laplace exponent \footnote{The proof is provided in Appendix A.4 in the supplementary material.} 
	
	\begin{equation}
	\label{Continous_stable}
	{\mathrm{\Phi }}_{V^{\left(\tau \right)}}\left(s\right)=s^{2^{-\tau }}{\left(2B\right)}^{1-2^{-\tau }},\,\, s>0.
	\end{equation}
	
	\noindent Thus, $V^{\left(0\right)}\left(t\right)=t$ , and for every $\tau>0$, $V^{\left(\tau \right)}\left(t\right)$, $t\ge0$ is an $\alpha$-stable subordinator with stable index $\alpha =2^{-\tau }$. We call $V^{\left(\tau \right)}\left(t\right)$, $t\ge0$ a \textit{$\tau$-compounded L\'{e}vy-stable subordinator} with  scale-intensity $B>0$. Thus, every $\alpha$-stable subordinator is a $\tau$-compounded L$\mathrm{\textrm{\'{e}}}$vy-stable subordinator with $\tau =-\frac{{ln \left(\alpha \right)\ }}{ln2}$, $\alpha \in \left(0,1\right)$.

	Consider next a log-price process $L^{\left(n\right)}_t=lnS_t$, $t\ge0,n=2,3,..$  of the form
	\begin{equation}
	\label{log_price_n}
	L^{\left(n\right)}_t=L^{\left(n\right)}_0+\mu t+\sum^n_{k=1}{{\gamma }_kV^{\left(k\right)}\left(t\right)}+\sigma B_{T(V^{\left(n\right)}\left(t\right))},t\ge 0,
	\end{equation}
	
	\noindent where $\mu\in R$, ${\gamma }_k\in R$, $k=1,2,..,$ and $\sigma>0$. Denote  \[{\mathrm{\Lambda }}^{\left(\mathrm{n}\right)}\mathrm{:}=L^{\left(n\right)}_1-L^{\left(n\right)}_0=\mu +\sum^n_{k=1}{{\gamma }_kV^{\left(n\right)}\left(1\right)}+\sigma B_{T\left(V^{\left(n\right)}\left(1\right)\right)}.\] Then, the Ch.f. of ${\mathrm{\Lambda }}^{\left(\mathrm{n}\right)}$, $n=2,3,\dots$ is given by \footnote{The proof is provided in Appendix A.5 in the supplementary material.} 
	
	\begin{equation}
	\label{chf_of_continous_stable}
	\footnotesize
	\begin{array}{cl}
	{\varphi }_{{\mathrm{\Lambda }}^{\left(\mathrm{n}\right)}}\left(v\right)=\mathbb{E}e^{iv{\mathrm{\Lambda }}^{\left(\mathrm{n}\right)}}=\\
	
	{\mathrm{exp}\left\{iv\mu-\sqrt{-2b_1\left(iv{\gamma }_1-\sqrt{-2b_2\left(\dots \sqrt{-2b_{n-1}\left(iv{\gamma }_{n-1}-\sqrt{-2b_n\left(iv{\gamma }_n-\frac{1}{2}v^2{\sigma }^2\right)}\right)}\dots \right)}\right)}\right\}}.
	\end{array}
	\end{equation}
	\normalsize
	We call $L^{\left(n\right)}_t$, $t\ge0$ with probability law determined by \eqref{chf_of_continous_stable}, a \textit{normal-compound(n)-stable log-price process}.
	Suppose $\mu=0$, then the characteristic exponent of ${\mathrm{\Lambda }}^{(n)}$ has the following recursive representation
	\[{\mathrm{\Psi }}_{{\mathrm{\Lambda }}^{\left(\mathrm{n}\right)}}\left(v\right)=-ln{\varphi }_{{\mathrm{\Lambda }}^{\left(\mathrm{n}\right)}}\left(v\right)=\sqrt{-2b_n\left(iv\gamma_n -{\mathrm{\Psi }}_{{\mathrm{\Lambda }}^{\left(\mathrm{n-1}\right)}}\left(v\right)\right)}.\]

	\subsection{Double-Gamma subordinator and Variance-Double-Gamma process}
	
	\noindent Here we consider the case when the L\'{e}vy subordinators \footnote{See \cite{Schoutens:2003} and \cite{Applebaum:2009}.} $T\left(t\right)$, $t\ge0$ and $U\left(t\right)$, $t\ge 0$ are gamma processes; that is,  $T\left(1\right)\sim Gamma\left({\alpha }_T,{\lambda }_T\right)$, \footnote{Here, $\sim$ stands for equal in distribution between two random variables or two stochastic processes.} ${\alpha}_T>0$, ${\lambda }_T\mathrm{>}0$ with pdf 
	\begin{equation}
	\label{Gamma_dis}
	f_{T\left(1\right)}\left(x\right)=\frac{{\lambda }^{{\alpha }_T}_T}{\mathrm{\Gamma }\left({\alpha }_T\right)}x^{{\alpha }_T-1}e^{-{\lambda }_Tx},\,\,x\ge0,
	\end{equation}
	and moment-generating function (MGF) 
	\begin{equation}
	\label{gamma_mgf}
	M_{T\left(1\right)}\left(v\right)={\left(1-\frac{v}{{\lambda }_T}\right)}^{-{\alpha }_T},v<{\lambda }_T,
	\end{equation}
	and $U\left(1\right)\sim Gamma\left({\alpha }_U,{\lambda }_U\right)$. We refer to $V\left(t\right)=T\left(U\left(t\right)\right),\,t\ge0$ as the \textit{double-gamma subordinator}. 
	
	The pdf and MGF of $\ V\left(1\right)=T\left(U\left(1\right)\right)$ are given by \footnote{The proof is provided in Appendix A.6 in the supplementary material.} 
	
	\begin{equation}
	\label{ pdf_gamma_subordinator}
	f_{V(1)}\left(x\right)=e^{-{\lambda }_Tx}\frac{{\lambda }^{{\alpha }_U}_U}{\mathrm{\Gamma }({\alpha }_U)}\int^{\infty }_0{\frac{{\lambda }^{{\alpha }_Tu}_T}{\mathrm{\Gamma }({\alpha }_Tu)}}x^{{\alpha }_Tu-1}u^{{\alpha }_U-1}e^{-{\lambda }_Uu}du,
	\end{equation}
	and
	\begin{equation}
	\label{MGF_gamma_subordinator}
	M_{V\left(1\right)}\left(v\right)={\left(1+\frac{{\alpha }_T}{{\lambda }_U}{\mathrm{ln} \left(1-\frac{v}{{\lambda }_T}\right)\ }\right)}^{-{\alpha }_U},
	\end{equation}
	for $0<v<{\lambda}_T\left(1-{\mathrm{exp}\left(\-\frac{{\lambda}_U}{{\alpha }_T}\right)}\right)$. Thus, $V\left(1\right)$ has a finite exponential moment $\mathbb{E}e^{v V\left(1\right)}$, for every $v\in \left(0,{\lambda }_T\left(1-{\mathrm{exp}\left(\-\frac{{\lambda }_U}{{\alpha }_T}\right)}\right)\right)$. From the representation of the MGF, we determined all four moments of $V(1)$. The mean of $V(1)$ is given by
	
	\begin{equation}
	\label{mean_gamm_sub}
	\mathbb{E}\left(V(1)\right)=\frac{{\alpha }_T}{{\lambda }_T}\frac{{\alpha }_U}{{\lambda }_U}=\mathbb{E}\left(T\left(1\right)\right)\mathbb{E}\left(U(1)\right).
	\end{equation}
	For the variance of $V\left(1\right)$ we have
	\begin{equation}
	\label{var_gamma_sub}
	var\left(V\left(1\right)\right)=\frac{{\alpha }_T}{{\lambda }^2_T}\frac{{\alpha }_U}{{\lambda }^2_U}\left({\alpha }_T+{\lambda }_U\right)=var\left(T\left(1\right)\right)var\left(U\left(1\right)\right)\left({\alpha }_T+{\lambda }_U\right),
	\end{equation}
	and the skewness of $V(1)$ is
	\begin{equation}
	\label{Skewness_gamma_sub}
	\begin{array}{ccc}
	Skewness\left[V(1)\right]=\frac{\mathbb{E}{\left[V\left(1\right)-\mathbb{E}V\left(1\right)\right]}^3}{{\left[var(V(1)\right]}^{\frac{3}{2}}}=$ $\frac{2}{\sqrt{{\alpha }_U}}\frac{1+\frac{3}{2}\frac{{\lambda }_U}{{\alpha }_T}+{\left(\frac{{\lambda }_U}{{\alpha }_T}\right)}^2}{{\left(1+\frac{{\lambda }_U}{{\alpha }_T}\right)}^{\frac{3}{2}}}
	\\ \ge \frac{2}{\sqrt{{\alpha }_U}}=$ $ Skewness\left[U\left(1\right)\right].
	\end{array}
	\end{equation}
	The equality is reached for $\frac{{\lambda }_U}{{\alpha }_T}\downarrow0$, \footnote{If together with $\frac{{\lambda }_U}{{\alpha }_T}\downarrow 0$, we also require that $var\left(V\left(1\right)\right)<\infty$, then $\frac{{\lambda }_U}{{\alpha }_T}\downarrow 0$.} implies ${\lambda }_U\downarrow 0$. Finally, the excess kurtosis of $T\left(U\left(1\right)\right)$ is given by
	\begin{equation}
	\label{excess_kurtosis_gamm_sub}
	\begin{array}{cc}
	ExcessKurtosis\left(V(1)\right)= \frac{\mathbb{E}{\left[V\left(1\right)-\mathbb{E}V\left(1\right)\right]}^4}{{\left[var(V(1)\right]}^2}-3
	= \frac{6}{{\alpha }_U}\frac{1+2\frac{{\lambda }_U}{{\alpha }_T}+\frac{11}{6}{\left(\frac{{\lambda }_U}{{\alpha }_T}\right)}^2+{\left(\frac{{\lambda }_U}{{\alpha }_T}\right)}^3}{{\left(1+\frac{{\lambda }_U}{{\alpha }_T}\right)}^2} \\ \ge \frac{6}{{\alpha }_U}=
	ExcessKurtosis\left(U\left(1\right)\right),
	\end{array}
	\end{equation}
	and the equality is reached for $\frac{{\lambda }_U}{{\alpha }_T}\downarrow0$.
	
	We now study the distributional characteristic of the \textit{variance-double-gamma process} 
	\begin{equation}
	\label{Variance-Double-Gamma}
	L_t=L_0+\mu t+\gamma U\left(t\right)+\rho V\left(t\right)+\sigma B_{V\left(t\right)}, \,\,t\ge0.
	\end{equation} 
	Because $L_t$, $t\ge0$ is a L\'{e}vy process, its distribution is determined by the unit increment
	\noindent $\mathrm{\Lambda }=L_1-L_0=\mu +\gamma U\left(1\right)+\rho V()+\sigma B_{V(1)}$ . We shall call $\mathrm{\ }\mathrm{\Lambda }$-distribution the \textit{variance-gamma-gamma distribution}. The pdf of $\mathrm{\Lambda }$ is given by \footnote{The proof is provided in Appendix A.7 in the supplementary material.}
	\begin{equation}
	\label{pdf_VGGD}
	\begin{array}{l}
	f_{\mathrm{\Lambda }}\left(x\right)=\frac{1}{\sqrt{2\pi }}\frac{{\lambda }^{{\alpha }_U}_U}{\mathrm{\Gamma }({\alpha }_U)}\int^{\infty }_0{\left(\int^{\infty }_0{e^{-\ \frac{{\left(\mathrm{x-}\mu -\gamma u-\rho y\right)}^2}{2{\sigma }^2y}}y^{{\alpha }_{Tu}-1}e^{-{\lambda }_Ty}dy}\right)\frac{{\lambda }^{{\alpha }_Tu}_T}{\mathrm{\Gamma }({\alpha }_{Tu})}}u^{{\alpha }_U-1}e^{-{\lambda }_Uu}du.
	\end{array}
	\end{equation}
	The expression for the pdf $f_{\mathrm{\Lambda }}\left(x\right)$, $x\in R$ is computationally intractable in view of the two integrals in the formula. We prefer to work with the MGF, $M_{\mathrm{\Lambda }}\left(v\right)=\mathbb{E}e^{v\mathrm{\Lambda }}$, $v>0$ which has the form
	\begin{equation}
	\label{mgf_VGGD}
	\begin{array}{ll}
	{ M}_{\mathrm{\Lambda }}\left(v\right)={\mathrm{exp} \left\{\mu v-{\alpha }_U{\mathrm{ln} \left[1-\frac{\gamma }{{\lambda }_U}v+\frac{{\alpha }_T}{{\lambda }_U}ln\left(1-\frac{\rho }{{\lambda }_T}v-\frac{{\sigma }^2}{2{\lambda }_T}v^2\right)\right]\ }\right\}}.
	\end{array}
	\end{equation}
	In \eqref{mgf_VGGD} we require that $0<v<\frac{\sqrt{{\rho}^2+2{\lambda }_T{\sigma }^2}-\rho }{{\sigma }^2}$ and   ${\lambda}_U+{\alpha}_Tln\left(1-\frac{1}{{\lambda}_T}\left(v\rho +\frac{1}{2}v^2{\sigma}^2\right)\right)\\-v\gamma >0$, which should be fulfilled when $v\downarrow 0.$ Given the representation \eqref{mgf_VGGD} we determine the four moments of $\mathrm{\Lambda }\mathrm{.}$ For the mean of $\mathrm{\Lambda }\mathrm{,}$ we have the following representation 
	\begin{equation}
	\label{mean_VGGD}
	\mathbb{E}\mathrm{\Lambda }\mathrm{=}\mu +\frac{{\alpha }_U}{{\lambda }_U}\gamma +\frac{{\alpha }_T}{{\lambda }_T}\frac{{\alpha }_U}{{\lambda }_U}\,\rho .
	\end{equation}

	The variance of $\mathrm{\Lambda }$ is given by
	\begin{equation}
	\label{var_VGGD}
	var\left(\mathrm{\Lambda }\right)\mathrm{=}\frac{{\alpha }_U}{{\lambda }^2_U}{\left(\frac{{\alpha }_T}{{\lambda }_T}\rho +\gamma \right)}^2+\frac{{\alpha }_U}{{\lambda }_U}\frac{{\alpha }_T}{{\lambda }_T}\left({\sigma }^2+\frac{{\rho }^2}{{\lambda }_T}\right).
	\end{equation}
	
	\noindent The skewness of $\mathrm{\Lambda }$ is 
	\begin{equation}
	\label{skew_VGGD}
	\begin{array}{lr}
	Skewness\left[\mathrm{\Lambda }\right]=\frac{\left({\alpha }_T\rho +{\lambda }_T\gamma \right)\left\{{\left({\alpha }_T\rho +{\lambda }_T\gamma \right)}^2+3{\lambda }_U{\alpha }_T{\lambda }_T\left({\rho }^2+{\sigma }^2{\lambda }_T\right)\right\}+{\lambda }^2_U{\alpha }_T\rho \left(2{\rho }^2+3{\sigma }^2{\lambda }_T\right)}{\sqrt{{\alpha }_U}{\left({\left({\alpha }_T\rho +{\lambda }_T\gamma \right)}^2+{\alpha }_T{\lambda }_U\left({\lambda }_T{\sigma }^2+{\rho }^2\right)\right)}^{\frac{3}{2}}}.
	\end{array}
	\end{equation}
	For the excess kurtosis of $\mathrm{\Lambda }\mathrm{,}$ we have
	\begin{equation}
	\label{EX_Kur_VGGD}
	\begin{array}{ll}
	\left(ExcessKurtosis\left(\mathrm{\Lambda }\right)\right)=\frac{\mathbb{E}{\left[\mathrm{\Lambda }-\mathbb{E}\mathrm{\Lambda }\right]}^4}{{\left[var\left(\mathrm{\Lambda }\right)\right]}^2}-3= \\
	\\
	\frac{\left( \begin{array}{c}
		6{\left({\alpha }_T\rho +\gamma {\lambda }_T\right)}^4+12{\lambda }_U{\alpha }_T\left({\rho }^2+{\sigma }^2{\lambda }_T\right){\left({\alpha }_T\rho +\gamma {\lambda }_T\right)}^2+ \\ 
		+4{\alpha }_T{\lambda }^2_U\rho \left(2{\rho }^2+3{\sigma }^2{\lambda }_T\right)\left({\alpha }_T\rho +\gamma {\lambda }_T\right)+ \\ 
		+3{\lambda }^3_U{\alpha }_T\ {\left(2{\rho }^2+{\sigma }^2{\lambda }_T\right)}^2+3{\lambda }^2_U{\alpha }^2_T{\lambda }^2_T{\left({\rho }^2+{\sigma }^2\right)}^2 \end{array}
		\right)}{{\alpha }_U{\left[{\left({\alpha }_T\rho +\gamma {\lambda }_T\right)}^2+{\lambda }_U{\alpha }_T\left({\sigma }^2{\lambda }_T+{\rho }^2\right)\right]}^2}.
	\end{array}
	\end{equation}
	\noindent Note that from \eqref{EX_Kur_VGGD} it follows that the excess kurtosis of  $\mathrm{\Lambda }$ can be negative if $\rho$ and $\gamma$ have opposite signs. In this case the distribution of $\mathrm{\Lambda }$ can become platykurtic. However, in the case where $\rho$ and $\gamma$ have the same sign, then the distribution of $\mathrm{\Lambda}$ is leptokurtic.
	
	Let us consider now the case of a compound subordination with multiple subordinators. Let $U^{\left(i\right)}\left(t\right)$, $t\ge0$, $i=1,\dots ,n$, $n\in \mathcal{N}=\left\{1,2,\dots \right\}$ be a sequence of independent gamma subordinators with $U^{\left(i\right)}\left(1\right)\sim Gamma\left({\alpha }_i,{\lambda }_i\right)$, and define $V^{\left(1\right)}\left(t\right)=U^{\left(1\right)}\left(t\right)$,   $V^{\left(i+1\right)}\left(t\right)=V^{\left(i\right)}\left(U^{\left(i+1\right)}\left(t\right)\right)$ for  $i=1,2,\dots ,n-1$. We shall use the notation  $V^{\left(n\right)}\left(t\right)=U^{\left(1\right)}\circ U^{\left(2\right)}\circ \dots \circ U^{\left(n\right)}\left(t\right),\ t\ge 0.$  Iteratively, we obtain the following representation for the MGF of $V^{\left(n\right)}\left(1\right)$ \footnote{The proof is provided in Appendix A.8  in the supplementary material.}, $n\in \mathcal{N}:$ 
	\begin{equation}
	\label{MGF_gamma_multi_subordinator}
	\begin{array}{ll}
	M_{V^{\left(n\right)}(1)}\left(v\right)={\left(1+\frac{{\alpha }_{n-1}}{{\lambda }_n}{\mathrm{ln} \left(1+\frac{{\alpha }_{n-2}}{{\lambda }_{n-1}}{\mathrm{ln} \dots {\mathrm{ln} \left(1+\frac{{\alpha }_1}{{\lambda }_2}{\mathrm{ln} \left(1-\frac{v}{{\lambda }_1}\right)}\right)}}\right)}\right)}^{-{\alpha }_n},
	\end{array}
	\end{equation}
	where $0<v<{\tau }_n<{\tau }_{n-1}$, and 
	\[{\tau }_n:={\lambda }_1\left(1-{\mathrm{exp} \left(-\frac{{\lambda }_2}{{\alpha }_1}\left(\dots \left(1-{\mathrm{exp} \left(-\frac{{\lambda }_{n-1}}{{\alpha }_{n-2}}\left(1-{exp \left(-\frac{{\lambda }_n}{{\alpha }_{n-1}}\right)}\right)\right)}\right)...\right)\right)}\right).\] 
	Note that for $n=2,3,\dots $ we have the recursive formula:
	\begin{equation}
	\label{recursive_MGF_Gamma}
	M_{V^{\left(n\right)}\left(1\right)}\left(v\right)={\left(1-\frac{1}{{\lambda }_n}lnM_{V^{\left(n-1\right)}\left(1\right)}\left(v\right)\right)}^{-{\alpha }_n},0<v<{\tau }_n.
	\end{equation}
	
	Formula \eqref{recursive_MGF_Gamma}, together with relations \eqref{mean_VGGD}-\eqref{EX_Kur_VGGD}, show that the probability mass of $V^{\left(n\right)}\left(1\right)$ (as $n\uparrow \infty )$ will either concentrate in $0$, as ${\tau }_n\downarrow 0$, or will escape to infinity, depending on the choice of $({\alpha }_n,{\lambda }_n)$ as $n\uparrow \infty$. There is no central limit theorem-type results for $V^{\left(n\right)}\left(1\right),n\uparrow \infty$, as there is no linear transformation of $V^{\left(n\right)}\left(1\right)$ leading to a proper distribution as a weak limit. It requires power-transformation of $V^{\left(n\right)}\left(1\right)$ to obtain non-trivial weak limits. However, those types of limiting results, while of potential academic interest, are beyond the scope of this paper. 
	
	Define the \textit{moment-generating exponent} of $V^{\left(n\right)}\left(t\right)$, $t\ge0$ as the cumulant-generating function of $V^{\left(n\right)}\left(1\right)$, ${\mathrm{K}}_{V^{\left(n\right)}}\left(v\right)=lnM_{V^{\left(n\right)}\left(1\right)}\left(v\right)=-{\alpha }_nln\left(1-\frac{1}{{\lambda }_n}{\mathrm{K}}_{V^{\left(n-1\right)}}\left(v\right)\right)$. Then the cumulants ${\mathrm{\kappaup }}_{j,n},j\in \mathcal{N}$, of  $V^{\left(n\right)}\left(1\right)$ are given by ${\mathrm{\kappaup }}_{j,n}={\left[\frac{{\partial }^j}{\partial v^j}{\mathrm{K}}_{V^{\left(n\right)}}\left(v\right)\right]}_{v=0}$, and $\mathbb{E}V^{\left(n\right)}\left(1\right)={\mathrm{\kappaup }}_{1,n}$, $var\left(V^{\left(n\right)}\left(1\right)\right)={\mathrm{\kappaup }}_{2,n}$,  ${Skewness}\left[V^{\left(n\right)}{(1)}\right]=\frac{\mathbb{E}{\left[V^{\left(n\right)}\left(1\right)-\mathbb{E}V^{\left(n\right)}\left(1\right)\right]}^3}{{\left[var(V^{\left(n\right)}(1)\right]}^{\frac{3}{2}}}=\frac{{\kappa }_{3,n}}{{\left({\kappa }_{2,n}\right)}^{\frac{3}{2}}}$, and \\${Excess Kurtosis} \left(V(1)\right)=\frac{\mathbb{E}{\left[V\left(1\right)-\mathbb{E}V\left(1\right)\right]}^4}{{\left[var(V(1)\right]}^2}-3=\frac{{\kappa }_{n,n}}{{\left({\kappa }_{2,n}\right)}^2}$.
	
	The following recursive formulas for ${\mathrm{\kappaup }}_{j,n},j=1,2,3$, and $4$ hold: \[{\mathrm{\kappaup }}_{1,n}=\frac{{\alpha }_n}{{\lambda }_n}{\mathrm{\kappaup }}_{1,n-1},\]
	\[{\mathrm{\kappaup }}_{2,n}=\frac{{\alpha }_n}{{\lambda }_n}\left(\frac{1}{{\lambda }_n}{\mathrm{\kappaup }}^2_{1,n}+{\mathrm{\kappaup }}_{2,n}\right),\] \[{\kappa }_{3,n}=\frac{{\alpha }_n}{{\lambda }_n}\left(\frac{2{\alpha }_n{\kappa }^3_{1,n-1}}{{\lambda }^2_n}+\frac{3{\alpha }_n{\kappa }_{1,n-1}{\kappa }_{2,n-1}}{{\lambda }^1_n}+{\kappa }_{3,n-1}\right),\]  \[ {\kappa }_{4,n}=\frac{{\alpha }_n}{{\lambda }_n}\left(\frac{6{\kappa }^4_{1,n-1}}{{\lambda }^3_n}+\frac{12{\kappa }^2_{1,n-1}{\kappa }_{2,n-1}}{{\lambda }^2_n}+\frac{3{\kappa }^2_{2,n-1}+4{\kappa }_{3,n-1}{\kappa }_{1,n-1}}{{\lambda }_n}+{\kappa }_{4,n-1}\right),\,\,n\ge 2\] 
	and ${\mathrm{\kappaup }}_{1,1}=\frac{{\alpha }_1}{{\lambda }_1},\,\, {\mathrm{\kappaup }}_{2,1}=\frac{{\alpha }_1}{{\lambda }^2_1},\,\,{\mathrm{\kappaup }}_{3,1}=\frac{2{\alpha }_1}{{\lambda }^3_1},\,\,{\mathrm{\kappaup }}_{4,1}=\frac{{6\alpha }_1}{{\lambda }^4_1}$.

	Let $U^{\left(i\right)}\left(t\right)$, $t\ge0$, $i=1,\dots ,n$, $n\in \mathcal{N}=\left\{1,2,\dots \right\}$ be a sequence of independent gamma subordinators with $U^{\left(i\right)}\left(1\right)\sim Gamma\left({\alpha }_i,{\lambda }_i\right)$, and define $V^{\left(1\right)}\left(t\right)=U^{\left(1\right)}\left(t\right)$,  $V^{\left(i+1\right)}\left(t\right)=V^{\left(i\right)}\left(U^{\left(i+1\right)}\left(t\right)\right)$ for  $i=1,2,\dots ,n-1$. Consider next a log-price process $L^{\left(n\right)}_t=lnS_t$, $t\ge0,n=2,3,..$  of the form
	\begin{equation}
	\label{Log_price_equation}
	L^{\left(n\right)}_t=L^{\left(n\right)}_0+\mu t+\sum^n_{k=1}{{\gamma }_k{\tilde{V}}^{\left(k\right)}\left(t\right)}+\sigma B_{V^{\left(n\right)}\left(t\right)},\,\,t\ge0,
	\end{equation}
	where $\mu\in R$, ${\gamma }_k\in R$, $\sigma>0$,  ${\tilde{V}}^{\left(k\right)}\left(t\right)=U^{\left(k\right)}\left(U^{\left(k-1\right)}\left(\dots \left(U^{\left(1\right)}\left(t\right)\right)\dots \right)\right)$, $k=1,\dots n$, $n\in \mathcal{N}$, and $U^{\left(i\right)}\left(t\right)$, $t\ge0,\ i=1,\dots ,n$ is a sequence of independent gamma subordinators with $U^{\left(i\right)}\left(1\right)\sim Gamma\left({\alpha }_i,{\lambda }_i\right)$.  Denote \textit{ }${\mathit{\Lambda}}^{\left(n\right)}:=L^{\left(n\right)}_1-L^{\left(n\right)}_0=\mu +\sum^n_{k=1}{{\gamma }_kV^{\left(k\right)}\left(1\right)}+\sigma B_{T\left(V^{\left(n\right)}\left(1\right)\right)}.\ $Then, the Ch.f. of ${\mathrm{\Lambda }}^{\left(\mathrm{n}\right)},n=2,3,\dots $ is given by \footnote{The proof is provided in Appendix A.9 in the supplementary material.}
	\begin{equation}
	\label{chf_multi_log_price}
	\begin{array}{cc}
	{\varphi }_{{\mathrm{\Lambda }}^{\left(\mathrm{n}\right)}}\left(v\right)=\\e^{iv\mu }{\left(1-iv\frac{{\gamma }_1}{{\lambda }_1}+\frac{{\alpha }_2}{{\lambda }_1}{ln \left(1-\dots -iv\frac{{\gamma }_{n-1}}{{\lambda }_{n-1}}+\frac{{\alpha }_n}{{\lambda }_{n-1}}ln\left(1-\ iv\frac{{\gamma }_n}{{\lambda }_n}+\frac{1}{2}v^2\frac{{\sigma }^2}{{\lambda }_n}\right)\dots \right) }\right)}^{-{\alpha }_1},\,\,v\in R.
	\end{array}
	\end{equation}

\subsection{Doubly-Inverse-Gaussian subordinator and Normal- Doubly- Inverse-Gaussian process.}

\noindent Here we  consider the case when the subordinators $T\left(t\right),t\ge 0$ and $U\left(t\right),t\ge 0$ are inverse Gaussian (IG) L\'{e}vy processes; that is,  $T\left(1\right)\sim IG\left({\lambda }_T,{\mu }_T\right)$, ${\lambda }_T\mathrm{\ >0}$, and ${\mu }_T>0$ with the pdf given by
\begin{equation}
\label{IG_dis}
f_{T\left(1\right)}\left(x\right)=\sqrt{\frac{{\lambda }_T}{2\pi x^3}}{\mathrm{exp} \left(-\frac{{\lambda }_T{\left(x-{\mu }_T\right)}^2}{2{\mu }^2_Tx}\right)\ },x\ge0,
\end{equation}
and $U\left(1\right)\sim IG\left({\lambda }_U, {\mu }_U\right)$. We refer to $V\left(t\right)=T\left(U\left(t\right)\right),t\ge 0$ as the \textit{double-inverse Gaussian subordinator}. We shall also consider the following two particular cases leading to single IG subordinators: $\left(i\right)\ {\mu }_T=1$, and ${\lambda }_T\uparrow \infty$, and thus $T\left(1\right)\to 1$ in ${\mathcal{L}}_2$ \footnote{A sequence $\{f_n\}$ of periodic, square-integrable functions is said	to converge in ${\mathcal{L}}_2$ to a function $f$ if the sequence of numbers $\int_{0}^{1}  \left| f_n(x)-f(x) \right|  ^2 dx$
	converges to 0.} and $\left(ii\right){\mu }_U=1$, and ${\lambda }_U\uparrow \infty $, and thus, $U\left(1\right)\to 1$ in ${\mathcal{L}}_2$.

The compound subordinator $V\left(t\right)=T(U\left(t\right)$, $t\ge0$ has pdf $f_{V\left(1\right)}\left(x\right)$ \footnote{The proof is provided in Appendix A.10 in the supplementary material.}  
\begin{equation}
\label{pdf_double_IG}
f_{V\left(1\right)}\left(x\right)=\frac{1}{2\pi }\sqrt{\frac{{\lambda }_T{\lambda }_U}{x^3}}\int^{\infty }_0{u^{-\frac{1}{2}}{\mathrm{exp} \left(-\frac{{\lambda }_T{\left(x-{\mu }_Tu\right)}^2}{2{\mu }^2_Tx}-\frac{{\lambda }_U{\left(u-{\mu }_T\right)}^2}{2{\mu }^2_Uu}\right)\ }}du,\,\, x>0.
\end{equation}

\noindent The MGF, $M_{V\left(1\right)}\left(v\right)=\mathbb{E}e^{vT\left(U\left(1\right)\right)}$, $v>0$  for $v\in \left(0,\frac{{\lambda }_T}{2{\mu }^2_T}\left[1-{\left(1-\frac{{\lambda }_U{\mu }_T}{2{\mu }^2_U{\lambda }_T}\right)}^2\right]\right)$ is \footnote{The proof is provided in Appendix A.10 in the supplementary material.}   
\begin{equation}
\label{MGF_double_LG}
M_{V(1)}\left(v\right)={\mathrm{exp} \left(\frac{{\lambda }_U}{{\mu }_U}\left(1-\sqrt{1-2\frac{{\mu }^2_U}{{\lambda }_U}\frac{{\lambda }_T}{{\mu }_T}\left(1-\sqrt{1-\frac{2{\mu }^2_T}{{\lambda }_T}v}\right)}\right)\right)},
\end{equation}
and the Ch.f of $V(1)$ is
\begin{equation}
\label{chf_double_LG}
\varphi_{V(1)}\left(v\right)={\mathrm{exp} \left(\frac{{\lambda }_U}{{\mu }_U}\left(1-\sqrt{1-2\frac{{\mu }^2_U}{{\lambda }_U}\frac{{\lambda }_T}{{\mu }_T}\left(1-\sqrt{1-\frac{2{\mu }^2_T}{{\lambda }_T}vi}\right)}\right)\right)}.
\end{equation}

To find the four central moments of $V\left( 1\right)$, we use the cumulant-generating function
${\mathrm{K}}_{\mathrm{V(1)}}\left(v\right)=ln{\ M}_{\mathrm{V(1)}}\left(v\right)$, and the cumulants ${\kappa }_n\mathrm{=}{\left[\frac{{\partial }^n}{\partial u^n}K_{V(1)}\left(u\right)\right]}_{u=0},\ n=1,2,3,4$.

\noindent Then, for the first two central moments of $V(1)$ we have 
\[ \mathbb{E}\mathrm{V(1)=}{\kappa }_1\mathrm{=}{\mu }_T{\mu }_U=\mathbb{E}T\left(1\right)\mathbb{E}U\left(1\right),\] and  \[Var\left(\mathrm{V(1)}\right)\mathrm{=}{\kappa }_2=\frac{{\mu }^3_U{\mu }^2_T}{{\lambda }_U}+\frac{{\mu }^3_T{\mu }_U}{{\lambda }_T}.\]

\noindent As $Var\left(T\left(1\right)\right)=\frac{{\mu }^3_T}{{\lambda }_T}$, and $Var\left(U\left(1\right)\right)=\frac{{\mu }^3_U}{{\lambda }_U}$, we have\\ \[Var\left(V\left(1\right)\right)=\left(Var\left(U\left(1\right)\right)\right){\left(\mathbb{E}\left(T(1\right)\right)}^2+\left(Var\left(T\left(1\right)\right)\mathbb{E}\left(U(1\right)\right).\]

\noindent Therefore, if  $\mathbb{E}\left( T\left( 1\right)\right)=1$, then $var\left(V\left(1\right)\right)>var\left(U\left(1\right)\right)$, and if $\mathbb{E}\left(U(1\right)=1$, then $var\left(V\left(1\right)\right)>var\left(T\left(1\right)\right)$. If $\mathbb{E}\left(T(1\right))={\mu }_T=1$ and ${\lambda }_T\uparrow \infty$, then $T\left(1\right)\to 1$ in ${\mathcal{L}}_2$-sense, and $var\left(V\left(1\right)\right)\to \frac{{\mu }^3_U}{{\lambda }_U}=var\left(U\left(1\right)\right)$. Similarly, if $\mathbb{E}\left(U(1\right))={\mu}_U=1$ and ${\lambda }_U\uparrow \infty $, then $U\left(1\right)\to 1$ in ${\mathcal{L}}_2$-sense and $Var\left(V\left(1\right)\right)\to =\frac{{\mu }^3_T}{{\lambda }_T}=Var\left(T\left(1\right)\right)$.

The skewness of $V(1)$ is given by 
\begin{equation}
\label{Skew_double_IG}
Skewness\left[\mathrm{V(1)}\right]={\kappa }_3{\kappa }^{-\frac{3}{2}}_2=3\frac{\frac{{\mu }^4_U}{{\lambda }^2_U}+\frac{{\mu }^2_U{\mu }_T}{{\lambda }_U{\lambda }_T}+\frac{{\mu }^2_T}{{\lambda }^2_T}}{{\mu }^{\frac{1}{2}}_U{\left(\frac{{\mu }^2_U}{{\lambda }_U}+\frac{{\mu }_T}{{\lambda }_T}\right)}^{\frac{3}{2}}}.
\end{equation}
If $\mathbb{E}\left(T(1\right))={\mu }_T=1$ and ${\lambda }_T\uparrow \infty$, then $T(1)\to 1$ in the ${\mathcal{L}}_2$-sense, and furthermore,  $Skewness\left[\mathrm{V\left(1 \right) }\right] \to 3\sqrt{\frac{{\mu }_U}{{\lambda }_U}}=Skewness\left[\mathrm{U(1)}\right].$ Similarly, if $\mathbb{E}\left(U(1\right))={\mu }_U=1$, and ${\lambda }_U\uparrow \infty $, then $Skewness\left[\mathrm{V}\left(\mathrm{1}\right)\right]\to 3\sqrt{\frac{{\mu }_T}{{\lambda }_T}}=Skewness\left[\mathrm{T}\left(\mathrm{1}\right)\right]$. 

Consider next the case when $\mathbb{E}\mathrm{T}\left(\mathrm{1}\right)=\mathbb{E}\mathrm{U}\left(\mathrm{1}\right)=VarT(1)=VarU(\left(1\right)=1$. Then ${Skewness}\left[\mathrm{T}\left(\mathrm{1}\right)\right]={Skewness}\left[\mathrm{U}\left(\mathrm{1}\right)\right]=3$, while ${Skewness}\left[\mathrm{V}\left(\mathrm{1}\right)\right]=3\frac{3}{2^{\frac{3}{2}}}=3.1819\dots$.

For the excess kurtosis of $\mathrm{V}\left(\mathrm{1}\right)$, we have the following expression
\begin{equation}
\label{Kortousis_double_IG}
\begin{array}{cc}
{ExcessKurtosis}\left(\mathrm{V(1)}\right)=\frac{{\kappa }_4}{{\kappa }^2_2}=
\frac{3\left[
	5 \left(\frac{\mu_T^2}{\lambda_U}    \right)^3+\,6 \left(\frac{\mu_U^4\,\mu_T}{\lambda_U^2\,\lambda_T}            \right)+\,5\left( \frac{\mu_U^2\mu_T^4}{\lambda_U\lambda_T^2}\right) +5\left( \frac{\mu_T^3}{\lambda_T^3}\right)                                   
	\right] }{\mu_T\left(\frac{\mu_U^2}{\lambda_U}+\frac{\mu_T}{\lambda_T} \right)^2 }.  
\end{array}
\end{equation}
If $\mathbb{E}\left(T(1\right))={\mu }_T=1$ and ${\lambda }_T\uparrow \infty $, then
\[{ExcessKurtosis}\left(\mathrm{V(1)}\right)\to 15\frac{{\mu }_U}{{\lambda }_U}={ExcessKurtosis}\left(\mathrm{U(1)}\right).\] 
If $\mathbb{E}\left(U(1\right))={\mu }_U=1$, and ${\lambda }_U\uparrow \infty$, then 
\[{ExcessKurtosis}\left(\mathrm{V(1)}\right)\to 15\frac{{\mu }_T}{{\lambda }_T}={ExcessKurtosis}\left(\mathrm{U(1)}\right).\] 

Now consider the case when $\mathbb{E}\mathrm{T}\left(\mathrm{1}\right)=\ \mathbb{E}\mathrm{U}\left(\mathrm{1}\right)=VarT(1)=VarU(\left(1\right)=1$. Then ${ExcessKurtosis}\left[\mathrm{T}\left(\mathrm{1}\right)\right]={ExcessKurtosis}\left[\mathrm{U}\left(\mathrm{1}\right)\right]=15$, while ${ExcessKurtosis}\left[\mathrm{V}\left(\mathrm{1}\right)\right]=15.75.$ 

Now let's study the distribution of the \textit{normal compound inverse Gaussian log-price process}, $L_t=lnS_t$, $t\ge0$, given by 
\[L_t=L_0+\mu t+\gamma U\left(t\right)+\rho T\left(U\left(t\right)\right)+\sigma B_{T(U\left(t\right))},\,t\ge0,\,\mu \in R,\,\gamma \in R,\,\rho \in R,\,\sigma >0,\] 
where the triplet $\left(B_s,T\left(s\right),U\left(s\right),s\ge0\right)$, $U\left(1\right)\sim \text{L\textrm{\'{e}}vy-inverse Gaussian}\left(\mu_U,\lambda_U\right)$,  $T\left(1\right)\sim \text{L\textrm{\'{e}}vy-inverse Gaussian}\left(\mu_T,\lambda_T\right)$  are independent processes generating stochastic basis  $(\mathrm{\Omega},\mathcal{F}, \mathbb{F}\\=({\mathcal{F}}_t,t\ge0),\mathbb{P})$ representing the natural world. $B_s$, $s\ge0$ is a standard Brownian motion and $T\left(s\right)$, $U\left(s\right)$, $s\ge0$, $\left( T\left(0\right)=0,\, U\left(0\right)=0 \right)$  are L\'{e}vy subordinators. Denote $\mathrm{\Lambda }\mathrm{:}=L_1-L_0=\mu +\gamma U\left(1\right)+\rho V(1)+\sigma B_{V(1)}$. The pdf of $\mathrm{\Lambda }$ is given by \footnote{The proof is provided in Appendix A.11  in the supplementary material.}
\begin{equation}
\label{dist_NCIG_log_price}
\begin{array}{ll}
f_{\mathrm{\Lambda }}\left(x\right)=\frac{1}{4\pi^2}\sqrt{\lambda_T\lambda_U}\int_{0}^{\infty}\int_{0}^{\infty}\frac{1}{4t^{\frac{3}{2}}}
\exp \left(-\frac{x-\mu-\gamma u-\rho t}{2\sigma\sqrt{t}}-\frac{\lambda_T\left(t-u\mu_T \right)^2}{2ut\mu_T^2 }-\frac{\lambda_U\left(u-\mu_U \right) }{2u\mu_U^2}      \right),
\end{array}                                                                        
\end{equation}
and for the Ch.f., ${\varphi }_{\mathrm{\Lambda }}\left(v\right)=\mathbb{E}e^{iv\mathrm{\Lambda }}$, we have the following expression
\begin{equation}
\label{chf_double_IG_log_price}
\begin{array}{cc}
{\varphi }_{\mathrm{\Lambda }}\left(v\right)=\mathbb{E}e^{iv\mathrm{\Lambda }}=
e^{ \,\,iv\mu+\frac{\lambda_U}{\mu_U} \left[ 1-\sqrt{1-\frac{2\mu_U^2}{\lambda_U}\left(\frac{\lambda_T}{\mu_T}\left(  1-\sqrt{1-\frac{2\mu_T^2}{\lambda_T}\left( iv\rho-\frac{1}{2}v^2\sigma^2\right) }\right) +iv\gamma\right) 
	}        
	\right] 
}.
\end{array}
\end{equation}
\noindent The MGF, ${M}_{\mathrm{\Lambda }}\left(u\right)$, is obtained by  setting $u=\frac{v}{i}$, and thus is omitted.

Having the representation given by \eqref{chf_double_IG_log_price}, we can determine the mean and the variance of $\Lambda$ as follows
\begin{equation}
\label{mean_NCIG}
\mathbb{E}\mathrm{\Lambda }\mathrm{=}\mu +\mu_U \gamma+\mu_U \mu_T \rho .
\end{equation}

\noindent and for the variance of $\mathrm{\Lambda}$, we have the following expression
\begin{equation}
\label{var_NCIG}
Var\left(\mathrm{\Lambda }\right)=\mu_U\frac{\rho^2 \mu_T^3}{\lambda_T}+\sigma^2\mu_U \mu_T +\frac{\mu_U^3\left(\gamma+\rho \mu_T \right)^2 }{\lambda_U}.
\end{equation}
Finally, the skewness of $\mathrm{\Lambda }$ is given by
\begin{equation}
\label{skew_doubal_NICG}
{Skewness}\left[\Lambda\right]=\frac{\frac{3\rho^3\mu_T^5}{\lambda_T^2}+\frac{3\rho\sigma^2\mu_T^3}{\lambda_T}+\frac{3\mu_T^4\left(\gamma+\mu_T \right)^3 }{\lambda_U^2}+\frac{3\mu_U^2}{\lambda_U}\left(\frac{\rho^2\mu_T^3}{\lambda_T}+\sigma^2\mu_T \right) \left(\gamma+\rho\mu_T \right)}                                
{\left(\mu_U\left(\sigma^2\mu_T+\frac{\rho^2\mu_T^3}{\lambda_t}+\frac{\mu_U^3}{\lambda_U} \left(\gamma+\rho\mu_T \right)^2 
	\right) 
	\right)^{\frac{3}{2}} }.
\end{equation}

We now consider the case of compound subordination with multiple subordinators. Let $U^{\left(i\right)}\left(t\right)$, $t\ge0$, $i=1,\dots ,n$, $n\in \mathcal{N}=\left\{1,2,\dots \right\}$ be a sequence of independent IG subordinators with $U^{\left(i\right)}\left(1\right)\sim IG\left({\mu }_i,{\lambda }_i\right)$, and define $V^{\left(1\right)}\left(t\right)=U^{\left(1\right)}\left(t\right)$,   $V^{\left(i+1\right)}\left(t\right)=V^{\left(i\right)}\left(U^{\left(i+1\right)}\left(t\right)\right)$ for  $i=1,2,\dots ,n-1.$. We shall use the notation  $V^{\left(n\right)}\left(t\right)=U^{\left(1\right)}\circ U^{\left(2\right)}\circ \dots \circ U^{\left(n\right)}\left(t\right),\ t\ge 0.$  Iteratively, we obtain the following representation for the Ch.f of $V^{\left(n\right)}\left(1\right)$, $n\in \mathcal{N}$ \footnote{The proof is provided in Appendix A.11  in the supplementary material.}:
\begin{equation}
\label{MGF_IG_n_compound}
\begin{array}{ll}
{\varphi }_{V^{\left(n\right)}\left(1\right)}\left(v\right)=e^{\frac{\lambda_n}{\mu_n}\left(1-\sqrt{1- \frac{2\mu_n^2\lambda_{n-1}}{\lambda_n\mu_{n-1}}\left( 1-\sqrt{1-\frac{2\mu_{n-1}^2\lambda_{n-2}}{\lambda_{n-1}\mu_{n-2}}\left(.....\sqrt{1-\frac{2\mu_2^2\lambda_1}{\lambda_2\mu_1}\left( 1-\sqrt{1-\frac{2\mu_1^2}{\lambda_1}iv}\right) } \right)  }\right)} \right)}.
\end{array}
\end{equation}
Note that for $n=2,3,...$, we have the following recursive formula  
\begin{equation}
\label{MGF_CNIG_Rec}
{\varphi }_{V^{\left(n\right)}\left(1\right)}\left(v\right)=e^{\frac{\lambda_n}{\mu_n}\left(1-\sqrt{1-\frac{2\mu_n^2}{\lambda_n}\ln {\varphi }_{V^{\left(n-1\right)}\left(1\right)}\left(v\right)} \right) }.
\end{equation}

Next consider a log-price process $L_t^{(n)}=Ln S_t$ of the form \eqref{Log_price_equation} and again denote
\[ {\mathit{\Lambda}}^{\left(n\right)}:=L^{\left(n\right)}_1-L^{\left(n\right)}_0=\mu +\sum^n_{k=1}{{\gamma }_kV^{\left(k\right)}\left(1\right)}+\sigma B_{T\left(V^{\left(n\right)}\left(1\right)\right)}.\] 
Then the chf of ${\mathit{\Lambda}}^{\left(n\right)}, n=2,3,...$, will be obtained iteratively  by the following representation:
\footnotesize
\begin{equation}
\label{chf_CNIG_logprice}
\begin{array}{cc}

{\varphi }_{{\mathrm{\Lambda }}^{\left(\mathrm{n}\right)}}\left(v\right)=e^{-\frac{\lambda_n}{\mu_n}\left[
	1-\sqrt{1-\frac{2\mu_n^2\lambda_{n-1}}{\lambda_n\mu_{n-1}}\left(1-\sqrt{1-\frac{2\mu_{n-1}^2\lambda_{n-2}}{\lambda_{n-1}\mu_{n-2}} \left(1-...\sqrt{1-\frac{2\mu_2^2\lambda_1}{\lambda_2\mu_1}\left(1-\sqrt{1-\frac{2\mu_1^2}{\lambda_1}\left( iv\gamma_1-\frac{1}{2}v^2 \sigma^2   \right) 
				}+iv\gamma_2 \right) 
			} \right)... 
		} \right) +
		iv\gamma_n}	
	\right]+iva 
}.
\end{array}
\end{equation}
\normalsize

\section{Empirical Analysis}
In this section, we apply the models we proposed in this paper to estimate the returns of a broad-based market index, the S\&P 500 as measured by SPR S\&P 500 which is an exchange-traded index. We use market indices by the triplet  $\left(L_t,T\left(U\left(t \right) \right),U\left(t\right)\right)$, $t\ge 0$ where: $\left(i\right)\ L_t$, $t\ge0$ as a stochastic model for the SPDR S\&P 500  index;  $\left(ii\right)\ V\left(t\right)$, $t\ge0$ as the cumulative VIX (i.e., $V\left(t\right)$ represents the cumulative value of VIX in $\left[0,t\right]$) (CBOE volatility index), and $\left(iii\right)\,U\left(t\right)$ $t\ge0$, as the cumulative VVIX (CBOE volatility of volatility index ) (i.e., $U\left(t\right)$, $t\ge0$ represents the cumulative value of VVIX in $\left[0,t\right]$). The subordinator processes $T\left(t\right),t\ge 0$ and $U\left(t\right),t\ge 0$ are inverse Gaussian L\'{e}vy processes (i.e.,  $T\left(1\right)\sim IG\left({\lambda }_T,{\mu }_T\right)$, ${\lambda }_T\mathrm{\ >0}$, and ${\mu }_T>0$).

In the log-return model, $\mathrm{\Lambda_t }=\mu t +\gamma U\left(t\right)+\rho V(t)+\sigma B_{V(t)}$, conditional on $U(t)$ and $V(t)$, the variance of $\mathrm{\Lambda_t }$ is $V(t)$. Therefore, the conditional volatility of $\mathrm{\Lambda_t }$ is $\sqrt{V(t)}$. Since the VIX index is a measure of the stock market's volatility, in modeling the variance of $\mathrm{\Lambda_t}$, $V(t)$, we use the squared value of the VIX index (VIX$^2$). 

Similarly, in the $V(t)$ log-return model conditional on $U(t)$, the variance of $V(t)$ is $U(t)$, and thus, conditional volatility is $\sqrt{U(t)}$. Since the VVIX index measures the volatility of the price of the VIX index, to model the variance of $V(t)$, we apply the squared value of VVIX index (VVIX$^2$) as a representation of the variance.

We then proceed as follows. First, we fit IG distribution to daily VVIX$^2$ data and compare the fitted density by the empirical kernel density. The kernel density estimator $\hat{f}_n(x)$,  for estimating the density of $f(x)$ at point \textit{x} is defined as 

\begin{equation}
\label{Kernel_density}
\hat{f}_n\left( x\right) =\frac{1}{nh}\sum_{i=1}^{n}k\left(\frac{x_i-x}{n} \right), 
\end{equation} 
where $k\left( x\right) =\frac{1}{\sqrt{2 \pi}} e^{-\frac{x^2}{2}}$ is the Gaussian kernel \citep[see][]{Epanechnikov:1969}. The mean and shape parameters of IG fitted on daily VVIX$^2$ index data over the period from January 2007 until the end of March 2019 using maximum likelihood methods are summarized in Table~\ref{table:IG_fit_VVIX}.

\begin{figure}[ht!]
	\centering
	\includegraphics[width=.8\textwidth]{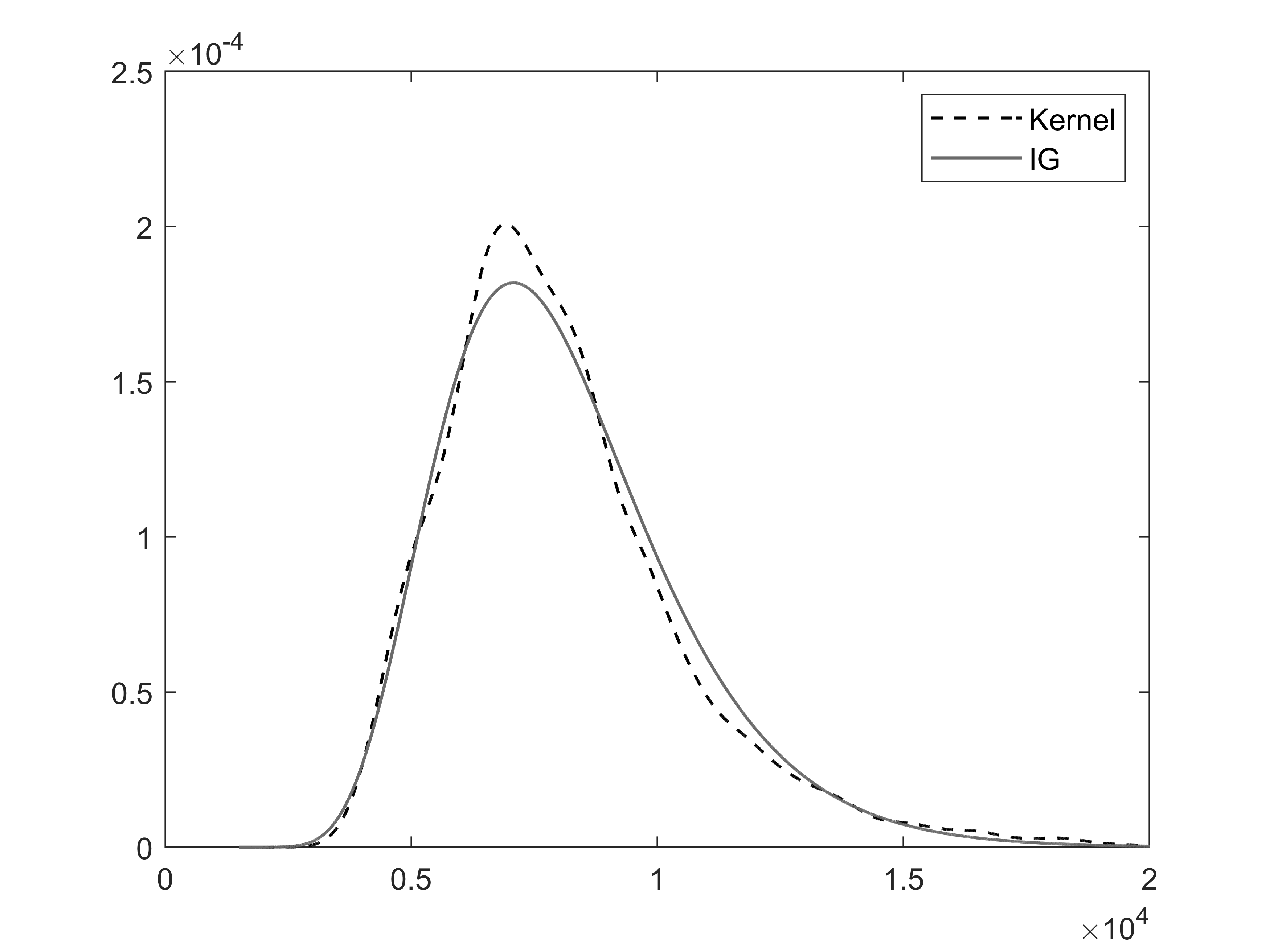}
	\caption{The IG fitted density via the kernel density of the daily VVIX$^2$ data.}
	\label{IG_fit_VVIX}
\end{figure}

Figure \ref{IG_fit_VVIX} shows the fitted IG distribution, corresponding to the empirical density for the daily VVIX$^2$ data. Our estimated model gives a good fit between the pdf and the empirical density of the data. The Kolmogorov--Smirnov test for goodness of fit testing verifies that the fitted IG is a good fit. The Kolmogorov--Smirnov value for the p--value$\left( \simeq1 \right)$  fails to reject the null hypothesis that the IG distributions are sufficient to describe the data.

In testing the  double subordinated model, we view $\left\lbrace  U(t), t\geq0\right\rbrace  $  as the stock-volatility intrinsic time (or volatility subordinator) and $\left\lbrace  V(t)=T(U(t)), t\geq0\right\rbrace$  as stock intrinsic time. 
In our model, $\left\lbrace  V(t)=T(U(t)),\right\rbrace$ is the variance of log-return process that is subordinated by IG volatility subordinator. Thus, $\left\lbrace  V(t)=T(U(t))\right\rbrace$ is a compound IG distribution (CIG) with four parameters. To estimate the model parameters, we fit the CIG distribution to daily VIX$^2$ index data. 
The method of modeling fitting via the empirical characteristic function (ECF) is applied to estimate the model parameters because of the difficulty in maximizing the likelihood function. We match the characteristic function derived from the CIG distribution to the ECF obtained from the daily VIX$^2$ data. The ECF procedure was first investigated by \cite{Paulson:1975} and recently by \cite{Yu:2003}. There is a one-to-one correspondence between the cumulative distribution function and the Ch.f because the pdf is the fast Fourier Transform (FFT) of the Ch.f.  Therefore, inference and estimation through the ECF are as efficient as the likelihood methods \citep[see][]{Yu:2003}. To estimate the model parameter, we minimized 
\begin{equation}
\label{emprical_ch_minimization}
h\left(r,x,\theta \right)=\int_{-\infty}^{\infty}\left( \frac{1}{n}\sum_{i=1}^{n} e^{i\theta x_i} -C\left( r,\theta\right)\right)  ^2 dr
\end{equation}  
where $C(r,\theta)$ is the Ch.f of $V(t)$ given by \eqref{chf_double_LG} . The daily VIX index data covering the period from January 1993 until the end of March 2019 consist of 6591 observations that we use to estimate the model's parameters. 

Because the CIG distribution has four parameters, the optimization method is more sensitive to the input of initial values and can simply fail to converge or converge to a local optimum. Here, the computational cost of estimating the four parameters model is high. The initial values are obtained from the method of moments estimation and additionally via instructed guesses. For any initial value we estimated the model parameters and consider the model as a good candidate to fit the data. 

To answer which model is the best in capturing the features of the data between the candidate density forecasts models, we first focus on the probability integral transforms (PIT) of the data in the evaluation of density models.
\cite{Diebold:1998} showed that a PIT time series  should be independent and identically distributed (iid) uniform  if the sequence of densities is correct. They proposed testing the specification of a density model by testing whether or not the transformed series is iid and uniform $(0,1)$. After evaluation of the density model, we selected the best model which was the one where the likelihood value is the largest.

\begin{figure}[t!]
	\centering
	\includegraphics[width=.8\textwidth]{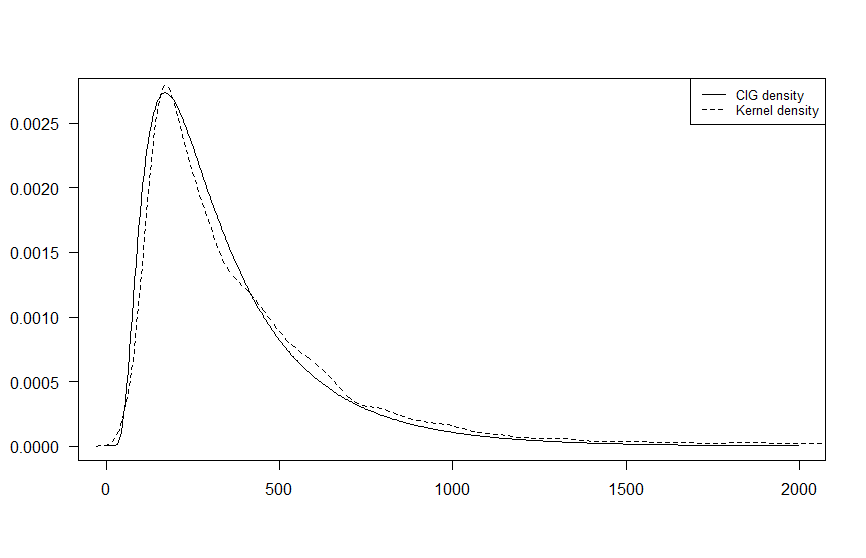}
	\caption{IG fitted CIG density via the kernel density of the daily VIX$^2$ data.}
	\label{CIG_fit_VIX}
\end{figure}

We implemented the FFT to calculate the pdf and then computed  the corresponding likelihood values. The estimated parameters of the best model are reported in Table~\ref{table:CIG_parameter}. The p-values$\left( \simeq1\right)$ of the \cite{Kolmogorov:1933} and \cite{Kuiper:1960} uniformity tests do not lead to rejecting the uniformity of the PIT.  Plotted in Figure~\ref{CIG_fit_VIX} is the CIG density with estimated parameters, corresponding to the empirical density of the daily VIX index. The figure reveals that our estimated model creates a good match between the pdf and the empirical density of the data.

As noted earlier, $V(t)$ is subordinated by an IG volatility subordinator, $U(t)$. From Table~\ref{table:CIG_parameter} it can be seen that this volatility subordinator exhibits an IG distribution with mean $\mu_U=172.7$ and shape parameter $\lambda_U=323.6$. Comparing these estimated parameters for the fitted IG distribution to the estimated parameters for the VVIX$^2$ index shown in Table~\ref{table:IG_fit_VVIX}, we see that there is a significant difference between the two models. 
This significant difference in mean and shape parameters obtained for the models is an indication that the VVIX index cannot be a proper volatility subordinator for the VIX index. In Table~\ref{table:VVIXcompare}, the mean, variance, skewness, and excess kurtosis for the volatility subordinator model and the IG distribution fitted to the VVIX$^2$ index are reported. 

In the case where we model the VIX index by using the VVIX index as the volatility subordinator, by comparing the skewness and excess kurtosis of the two models we can see again there is a significant difference in skewness and kurtosis for both models. This suggests that using the VVIX index as a measure of time change cannot contain all the information of stochastic volatility models; that is, the skewness and the fat-tail phenomenon of the VIX index are not properly captured by the VVIX index. Thus to have a proper model for the VIX index, the VIX's skewness and fat-tail phenomenon should be recovered by specifying a different volatility subordinator.

Next we investigate the distribution of $\mathrm{\Lambda_t }=\mu t +\gamma U\left(t\right)+\rho V(t)+\sigma B_{V(t)}$ as a stochastic model for the SPDR S\&P 500  log-return index by fitting a normal compound inverse Gaussian (NCIG) distribution to the data. $\mathrm{\Lambda_t }$ is a stochastic process with eight parameters, four of the parameters of $\mathrm{\Lambda_t }$ enter the model because of the intrinsic time change process. To estimate the parameters of the model, we use daily log-returns of the SPDR S\&P 500  index based on closing prices by implementing the Ch.f method. The database covers the period from January 1993 to March 2019 and includes 6591 observations collected from Yahoo Finance. As before, the optimization method is more sensitive to the input of initial values. The method of moments and instructed guess are used to obtain the initial values. We implemented the FFT to calculate the pdf and calculate the corresponding likelihood values. The best model to fit and explain the observed data is chosen as the one with the largest likelihood value. To evaluate the forecast density, we applied the PIT and inverse-normal-transform of the probability integral transform that should be iid standard normal as presented in \cite{BERKOWITZ:200}. 
In case of rejection of any tests, we changed the initial values and iterated the process. Finally, we calculated the likelihood value and selected the best model by comparing their likelihoods. The estimated parameters of the best model are summarized in Table~\ref{NCIG_parameters}. The p-values of the Kolmogorov--Smirnov (p-value$=1$) and Kuipers (p-value$=1$) uniformity tests do not lead to rejecting the uniformity of PIT. The performed adjusted Jarque--Bera test \citep[see][]{Urzua:1996} for the composite hypothesis of normality (p-value$=0.093$) fails to reject the null hypothesis of inverse-normal-transform that the data are normally distributed.

\begin{figure}[ht!]
	\centering
	\includegraphics[width=.8\textwidth]{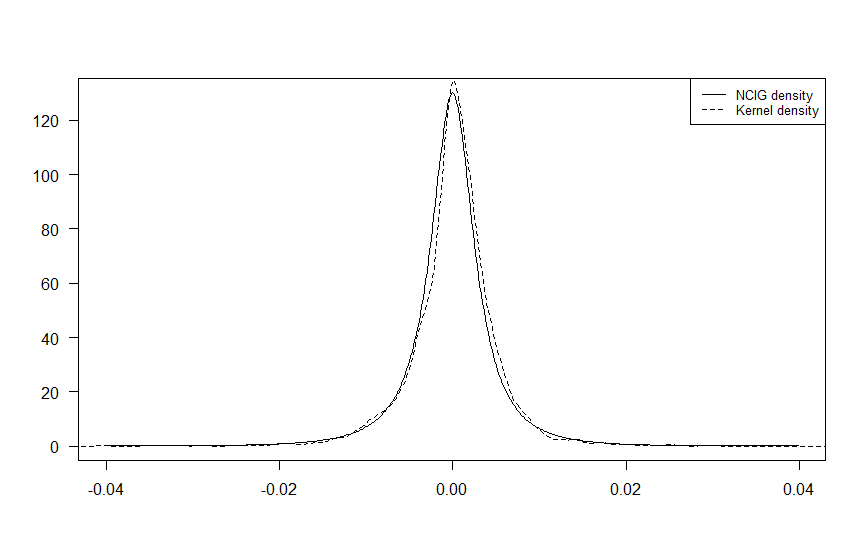}
	\caption{The NCIG density of log-return SPDR S\&P 500  via the kernel density.}
	\label{figure:SPY_fit}
\end{figure}

The model density estimates corresponding to the empirical density of the daily log-return SPDR S\&P 500  index are plotted in Figure~\ref{figure:SPY_fit}. The figure reveals that our estimated model creates a good match between the pdf and the empirical density of the data. 

There is a question as to whether the VIX index is a proper time change subordinator for the SPDR S\&P 500  log-return model. An intuitive way to answer the question is by comparing the skewness and excess kurtosis for the subordinator models in the SPDR S\&P 500  log-return  with the CIG distribution fitted to the VIX data. The first four standardized moments of both models are given in Table \ref{table:VIX_compare}. The results indicate that the model fitted on VIX$^2$ data exhibits a heavy tail in contrast to the subordinated model for the SPDR S\&P 500  log-return model.  Also, we observe that the skewness of the CIG model fitted to VIX$^2$ is more extreme than the time change subordinated model. Thus, it can be concluded that the VIX index is not a proper intrinsic time change for the SPDR S\&P 500  index. We see that an index with a thinner tail and slight positive skewness than the VIX index can improve the SPDR S\&P 500  log-return model.

Finally, we mention that for the SPDR S\&P 500-log return model given by~\eqref{log_return_eq}, the coefficient of the volatility subordinator, $\gamma$, is zero. This finding suggests that the VVIX index does not have too much influence directly in modeling the log-return of SPDR S\&P 500. This is because the VVIX is an indicator of the expected volatility of the VIX index, and VIX is not a proper time change subordinate in the model.

\section{Conclusion}
In this paper, we generalized the classical asset pricing model by replacing physical time in the well-known return model with multiple stochastic intrinsic times subordinator. This modification to the return model takes into account tail effects, one of the stylized facts known about stock return. We introduced the stock-volatility intrinsic time or volatility subordinator to the model to reflect the heavy-tail phenomena present in asset returns. This increased the number of parameters that are required to be estimated. The properties of the $\alpha$-stable, gamma and inverse Gaussian multiple subordinator models are described. We defined the normal double stable, variance double gamma processes, and normal double inverse Gaussian processes for modeling asset returns. Our empirical results suggest that the VIX and VVIX indexes are not the proper intrinsic time change and volatility subordinators for modeling the SPDR S\&P 500  log-return, respectively.


\begin{thebibliography}{}
	
	\bibitem[\protect\citeauthoryear{Abate and Whitt}{Abate and
		Whitt}{1999}]{Abate:1999}
	Abate, J. and W.~Whitt (1999).
	\newblock Laplace transforms of probability density functions with series
	representations.
	\newblock {\em The Operations Research Society of Japan\/}~{\em 42}, 268--285.
	
	\bibitem[\protect\citeauthoryear{Applebaum}{Applebaum}{2009}]{Applebaum:2009}
	Applebaum, D. (2009).
	\newblock {\em L\'{e}vy Processes and Stochastic Calculus}.
	\newblock Cambridge: Cambridge University Press.
	
	\bibitem[\protect\citeauthoryear{Berkowitz}{Berkowitz}{2001}]{BERKOWITZ:200}
	Berkowitz, J. (2001).
	\newblock Testing density forecasts, with applications to risk management.
	\newblock {\em Journal of Business and Economic Statistics\/}~{\em 19},
	465--474.
	
	\bibitem[\protect\citeauthoryear{Bochner}{Bochner}{1995}]{Bochner:1955}
	Bochner, S. (1995).
	\newblock {\em Harmonic Analysis and the Theory of Probability}.
	\newblock University of California Press: Berkeley and Los Angeles.
	
	\bibitem[\protect\citeauthoryear{Carr, Geman, and Madan}{Carr
		et~al.}{2003}]{Carr:2003}
	Carr, P., H.~Geman, and D.~Madan (2003).
	\newblock Stochastic volatility for {L}\'{e}vy processes.
	\newblock {\em Mathematical Finance\/}~{\em 13}, 345--382.
	
	\bibitem[\protect\citeauthoryear{Carr and Wu}{Carr and Wu}{2004}]{Carr:2004}
	Carr, P. and L.~Wu (2004).
	\newblock Time-changed {L}\'{e}vy processes and option pricing.
	\newblock {\em Financial Economics\/}~{\em 17}, 113--141.
	
	\bibitem[\protect\citeauthoryear{Carrasco and Kotchoni}{Carrasco and
		Kotchoni}{2017}]{Carrasco:2017}
	Carrasco, M. and R.~Kotchoni (2017).
	\newblock Efficient estimation using the characteristic function.
	\newblock {\em Econometric Theory\/}~{\em 33}, 479--526.
	
	\bibitem[\protect\citeauthoryear{Clark}{Clark}{1973}]{Clark:1973}
	Clark, P. (1973).
	\newblock A subordinated stochastic process model with fixed variance for
	speculative prices.
	\newblock {\em Econometrica\/}~{\em 41}, 135--156.
	
	\bibitem[\protect\citeauthoryear{Cont}{Cont}{2001}]{Cont:2001}
	Cont, R. (2001).
	\newblock Empirical properties of asset returns: Stylized facts and statistical
	issues.
	\newblock {\em Quantitative Finance\/}~{\em 1}, 223--236.
	
	\bibitem[\protect\citeauthoryear{Diebold, Gunther, and Tay}{Diebold
		et~al.}{1998}]{Diebold:1998}
	Diebold, F., T.~Gunther, and T.~Tay (1998).
	\newblock Evaluating density forecasts.
	\newblock {\em International Economic Review\/}~{\em 39}, 863--883.
	
	\bibitem[\protect\citeauthoryear{Epanechnikov}{Epanechnikov}{1969}]{Epanechnikov:1969}
	Epanechnikov, V. (1969).
	\newblock Non-parametric estimation of a multivariate probability density.
	\newblock {\em Theory of Probability and Its Applications\/}~{\em 14},
	153--158.
	
	\bibitem[\protect\citeauthoryear{Glasserman and Liu}{Glasserman and
		Liu}{2010}]{Glasserman:2010}
	Glasserman, P. and Z.~Liu (2010).
	\newblock Sensitivity estimates from characteristic functions.
	\newblock {\em Operations Research\/}~{\em 58}, 1611--1623.
	
	\bibitem[\protect\citeauthoryear{Hurst, Platen, and Rachev}{Hurst
		et~al.}{1997}]{Hurst:1997}
	Hurst, S., E.~Platen, and S.~Rachev (1997).
	\newblock Subordinated market index models: A comparison.
	\newblock {\em Financial Engineering and the Japanese Markets\/}~{\em 4},
	97--124.
	
	\bibitem[\protect\citeauthoryear{Jacod and Shiryaev}{Jacod and
		Shiryaev}{2005}]{Jacod:2003}
	Jacod, J. and A.~Shiryaev (2005).
	\newblock {\em Limit Theorems for Stochastic Processes}.
	\newblock Berlin: Springer Verlag.
	
	\bibitem[\protect\citeauthoryear{Kateregga, Mataramvura, and Taylor}{Kateregga
		et~al.}{2017}]{Kateregga:2017}
	Kateregga, M., S.~Mataramvura, and D.~Taylor (2017).
	\newblock Parameter estimation for stable distributions with application to
	commodity futures log-returns.
	\newblock {\em Cogent Economics \& Finance\/}~{\em 5}, 1--28.
	
	\bibitem[\protect\citeauthoryear{Klingler, Kim, Rachev, and Fabozzi}{Klingler
		et~al.}{2013}]{Klingler:2013}
	Klingler, S., Y.~Kim, S.~Rachev, and F.~Fabozzi (2013).
	\newblock Option pricing with time-changed {L}\'{e}vy processes.
	\newblock {\em Applied Financial Economics\/}~{\em 23\/}(15), 12--31.
	
	\bibitem[\protect\citeauthoryear{Kolmogorov}{Kolmogorov}{1933}]{Kolmogorov:1933}
	Kolmogorov, A. (1933).
	\newblock Sulla determinazione empirica di una legge di distribuzione.
	\newblock {\em G. Ist. Ital. Attuari\/}~{\em 4}, 83--91.
	
	\bibitem[\protect\citeauthoryear{Kuiper}{Kuiper}{1960}]{Kuiper:1960}
	Kuiper, N. (1960).
	\newblock Evaluating density forecasts.
	\newblock {\em International Economic Review\/}~{\em 63}, 38--47.
	
	\bibitem[\protect\citeauthoryear{Mandelbrot and Taylor}{Mandelbrot and
		Taylor}{1967}]{Mandelbrot:1967}
	Mandelbrot, B. and H.~Taylor (1967).
	\newblock On the distribution of stock price differences.
	\newblock {\em Operations Research\/}~{\em 15}, 1057--1062.
	
	\bibitem[\protect\citeauthoryear{Mnatsakanov and Sarkisian}{Mnatsakanov and
		Sarkisian}{2013}]{Mnatsakanov:2013}
	Mnatsakanov, R. and K.~Sarkisian (2013).
	\newblock A note on recovering the distributions from exponential moments.
	\newblock {\em Applied Mathematics and Computation\/}~{\em 219}, 8730--8737.
	
	\bibitem[\protect\citeauthoryear{Paulson, Holcomb, and Leitch}{Paulson
		et~al.}{1975}]{Paulson:1975}
	Paulson, A., E.~Holcomb, and R.~Leitch (1975).
	\newblock The estimation of the parameters of the stable laws.
	\newblock {\em Biometrika\/}~{\em 62}, 163--170.
	
	\bibitem[\protect\citeauthoryear{Prelec}{Prelec}{1998}]{Prelec:1998}
	Prelec, D. (1998).
	\newblock The probability weighting function.
	\newblock {\em Econometrica\/}~{\em 66}, 497--527.
	
	\bibitem[\protect\citeauthoryear{Rachev, Fabozzi, and Racheva-Iotova}{Rachev
		et~al.}{2017}]{Rachev:2017}
	Rachev, S., F.~Fabozzi, and B.~Racheva-Iotova (2017).
	\newblock Option pricing with greed and fear factor: The rational finance
	approach.
	\newblock {\em arXiv:1709.08134[q-fin.GN]\/}.
	
	\bibitem[\protect\citeauthoryear{Rachev, Menn, and Fabozzi}{Rachev
		et~al.}{2005}]{Rachev:2005}
	Rachev, S., C.~Menn, and F.~Fabozzi (2005).
	\newblock {\em Fat-Tailed and Skewed Asset Return Distributions}.
	\newblock Hoboken,NJ: John Wiley \& Sons.
	
	\bibitem[\protect\citeauthoryear{Samorodnitsky and Taqqu}{Samorodnitsky and
		Taqqu}{1994}]{Samorodnitsky:1994}
	Samorodnitsky, G. and M.~Taqqu (1994).
	\newblock {\em Stable Non-Gaussian Random Processes}.
	\newblock Boca Raton: Chapman \& Hall/CRC.
	
	\bibitem[\protect\citeauthoryear{Sato and Katok}{Sato and
		Katok}{1999}]{Sato:2002}
	Sato, K. and A.~Katok (1999).
	\newblock {\em L\'{e}vy Processes and Infinitely Divisible Distributions}.
	\newblock Cambridge University Press.
	
	\bibitem[\protect\citeauthoryear{Schoutens}{Schoutens}{2003}]{Schoutens:2003}
	Schoutens, W. (2003).
	\newblock {\em L\'{e}vy Processes in Finance: Pricing Financial Derivatives}.
	\newblock John Wiley \& Sons.
	
	\bibitem[\protect\citeauthoryear{Tagliani and Vel\'{a}sques}{Tagliani and
		Vel\'{a}sques}{2004}]{Tagliani:2004}
	Tagliani, A. and Y.~Vel\'{a}sques (2004).
	\newblock Inverse {L}aplace transform for heavy-tailed distributions.
	\newblock {\em Applied Mathematics and Computations\/}~{\em 150}, 337--345.
	
	\bibitem[\protect\citeauthoryear{Tsionas}{Tsionas}{2012}]{Tsionas:2012}
	Tsionas, E. (2012).
	\newblock Maximum likelihood estimation of stochastic frontier models by the
	{F}ourier transform.
	\newblock {\em Journal of Econometrics\/}~{\em 170}, 234--248.
	
	\bibitem[\protect\citeauthoryear{Tversky and Kahneman}{Tversky and
		Kahneman}{1992}]{Tversky:1992}
	Tversky, A. and D.~Kahneman (1992).
	\newblock Advances in prospect theory: Cumulative representation of
	uncertainty.
	\newblock {\em Risk and Uncertainty\/}~{\em 5}, 297--232.
	
	\bibitem[\protect\citeauthoryear{Urzua}{Urzua}{1996}]{Urzua:1996}
	Urzua, C. (1996).
	\newblock On the correct use of omnibus tests for normality.
	\newblock {\em Economics Letters\/}~{\em 53}, 247--251.
	
	\bibitem[\protect\citeauthoryear{Yu}{Yu}{2003}]{Yu:2003}
	Yu, J. (2003).
	\newblock Empirical characteristic function estimation and its applications.
	\newblock {\em Econometric Reviews\/}~{\em 23}, 93--123.
	
\end{thebibliography}
\bibliographystyle{chicago}

\clearpage
\begin{table}[]
	\centering
	\caption{Estimated parameters of IG fitted to the daily VVIX$^2$ data.}
	\begin{tabular}{@{}cc@{}}
		\toprule
		$\,\,\,\,\,\,\,\,\mu_U$ &\,\,\,\,\,\,\,\,\,\,\,\, $\lambda_U$ \\ \midrule
		\,\,\,\,\,\,\,8096.84 & \,\,\,\,\,\,\,\,\,\,\,\,\,\,90189.7     \\ \bottomrule
	\end{tabular}
	\label{table:IG_fit_VVIX}
\end{table}

\begin{table}[]
	\begin{center}
		\caption{The estimated Parameters of  CIG distribution fitted to daily VIX$^2$}
		\begin{tabular}{@{}lllll@{}}
			\toprule
			Parameters & \,\,\,\,\,$\lambda_U$ \,\,\,\,\, & \,\,\,$\,\,\mu_U$   \,\,\,\,\, &\,\,\, $\lambda_T$ \,\,\,\,\,   & \,\,\,$\mu_T$\,\,\,\,\, \\ \midrule
			Estimates & 323.6 &\, 172.7 & \,\,\,20.1 & 2.05                    \\ \bottomrule
		\end{tabular}
		\label{table:CIG_parameter}	
	\end{center}	
\end{table}

\begin{table}[]
	\begin{center}
		\caption{Mean, variance, skewness, and excess kurtosis of IG distribution, in Volatility Subordinator model and IG model fitted to VVIX$^2$  }
		\begin{tabular}{@{}lcc@{}}
			\toprule
			Model &\,\,\,\,\,\,VVIX$^2$ model   & \,\,\,\,\,
			Volatility Subordinator model\\ \midrule
			Mean     & 8096.8                                      & 172.7                                 \\
			Variance & 5885600                                     & 15917                                  \\
			Skewness & 0.8989                                      & 2.1916                                 \\
			Excess Kurtosis & -1.6534                                    & 5.0053                                 \\ \bottomrule
		\end{tabular}
		\label{table:VVIXcompare}
	\end{center}
\end{table}

\begin{table}[]
	\centering
	\caption{The estimated parameters of  NCIG distribution fitted to daily SPDR S\&P 500  log-returns}
	\begin{tabular}{@{}cccccccc@{}}
		\toprule
		$\lambda_U$          &\,\,\,\,\,\,\,\,\,\, $\mu_U$        & \,\,\,\,\,\,\,\,\,\,$\lambda_T$         &\,\,\,\,\,\,\,\,\,\, $\mu_T$       &\,\,\,\, \,\,\,\,\,\,$\mu$        & \,\,\,\,\,\,\,\,\,\,$\gamma$    & \,\,\,\,\,\,\,\,\,\,$\rho$         &\,\,\,\,\,\, $\sigma$       \\ \midrule
		\,\,\,\,\,\,\,\,\,\, 17.66 & \,\,\,\,\,\,\,\,\,\,0.0035 & \,\,\,\,\,\,\,\,\,\,12.54 &\,\,\,\,\,\,\,\,\,\, 0.122 & \,\,\,\,\,\,\,\,\,\,0.0 &  \,\,\,\,\,\,\,\,\,\,0.0 & \,\,\,\,\,\,\,\,\,\,-0.281 &\,\,\,\,\,\,\,\,\,\, 0.252 \\ \bottomrule
	\end{tabular}
	\label{NCIG_parameters}
\end{table}

\begin{table}[t]
	\begin{center}
		\caption{Mean, variance, skewness, and excess kurtosis of CIG distribution in stock intrinsic time subordinator model and CIG model fitted to VIX$^2$.  }
		\begin{tabular}{@{}lcc@{}}
			\toprule
			Model &  CIG model fitted to VIX index & \,\,\,\,\,\,  SPDR time subordinator model\\ \midrule
			Mean     & 354.03                                    & 0.2167                                 \\
			Variance & 66966                                     & 0.0025                                 \\
			Skewness & 2.191                                     & 0.6313                                 \\
			Kurtosis & 7.998                                     & 0.6707                               \\ \bottomrule
		\end{tabular}
		\label{table:VIX_compare}
	\end{center}
\end{table}

\clearpage


\section*{Appendix}

\begin{appendix}
	\indent	\appendix{\textbf{A.1: Characteristic function of ${\alpha}_T$-stable process}}
	\label{appendix:i}
	
	\noindent If $B_{T(t)}$ is an ${\alpha}_T$-stable motion with unit increment $B_{T(1)}$, then the characteristic function (Ch.f.) of $B_{T(t)}$ is given by\\
	\noindent${\varphi}_{B_{T\left(1\right)}}(u)={\mathbb{E}}_{T(1)=v}{\mathrm{exp}\left\{-v\frac{u^2}{2}\right\}}=\mathbb{E}{\mathrm{exp} \left\{-T(1)\frac{u^2}{2}\right\}}={\mathfrak{L}}_{T\left(1\right)}\left(\frac{u^2}{2}\right)={\mathrm{exp}\left(-{\left(\frac{{\delta }_T}{2}\right)}^{\frac{{\alpha }_T}{2}}u^{{\alpha }_T}\right)}$.\\
\end{appendix}

\begin{appendix}
	\appendix{\textbf{A.2: Laplace exponent of double ${\alpha}_T$-stable subordinator}}
	\label{appendix:ii}
	
	\noindent The Laplace exponent of the compound subordinator $V\left(t\right)=T\left(U\left(t\right)\right),t\ge 0$ where
	$U\left(1\right)\sim \text{L\textrm{\'{e}}vy-stable}\left(b_U\right)$  and $T\left(1\right)\sim\text{L\textrm{\'{e}}vy-stable}\left(b_T\right)$ 
	are independent processes, is given by\\  
	\noindent${\mathrm{\Phi}}_V\left(s\right)=-{\mathrm{ln} \left({\mathbb{E}}_{U\left(1\right)=u}\mathbb{E}e^{-sT\left(u\right)}\right)\ }=-{\mathrm{ln} \left({\mathbb{E}}_{U\left(1\right)=u}{\left({\mathrm{exp} \left(-{\mathit{\Phi}}_T\left(s\right)\right)\ }\right)}^u\right)\ }
	=-{\mathrm{ln} \left(\mathbb{E}{\mathrm{exp} \left(-U\left(1\right){\mathit{\Phi}}_T\left(s\right)\right)}\right)}\\={\mathit{\Phi}}_U\left({\mathit{\Phi}}_T\left(s\right)\right)
	={({\delta }_U{\mathrm{\Phi }}_T\left(s\right))}^{\frac{{\alpha }_U}{2}}
	={\left({\delta }_U{\left({\delta }_Ts\right)}^{\frac{{\alpha }_T}{2}}\right)}^{\frac{{\alpha }_U}{2}}={\delta }^{\frac{{\alpha }_U}{2}}_U{\left({\delta }_Ts\right)}^{\frac{{\alpha }_T}{2}\frac{{\alpha }_U}{2}}.$ \\
	
\end{appendix}

\begin{appendix}
	\appendix{\textbf{A.3: Ch.f. of normal-double-stable log-price process}}
	\label{appendix:iii}
	
	\noindent Let $L_t$ be a  \textit{normal-compound-stable log-price process} 
	\[L_t=L_0+\mu t+\gamma U\left(t\right)+\rho T\left(U\left(t\right)\right)+\sigma B_{T(U\left(t\right))},\,t\ge0,\,\mu\in R,\,\gamma \in R,\,\rho \in R,\,\sigma >0,\] 
	where the triplet $\left(B_s, T\left(s\right), U\left(s\right), s\ge0\right)$, 
	$U\left(1\right)\sim \text{L\textrm{\'{e}}vy-stable}\left(b_U\right)$ , $T\left(1\right)\sim\text{L\textrm{\'{e}}vy-stable}\left(b_T\right)$ 
	are independent processes,
	and $B_s$, $s\ge0$ is a standard Brownian motion, and $T\left(s\right)$, $U\left(s\right)$, $s\ge0$, $\left( T\left(0\right)=0, U\left(0\right)=0 \right)$  are L\'{e}vy subordinators, Denote $\mathrm{\Lambda }\mathrm{:}=L_1-L_0=\mu +\gamma U\left(1\right)+\rho V(1)+\sigma B_{V(1)}$, then the Ch.f. of $\mathrm{\Lambda }$ is given by\\
	\indent ${\varphi }_{{\mathrm{\Lambda }}_{\mathrm{1}}}\left(v\right)=e^{-{\mathrm{\Psi }}_{\mathrm{\Lambda }}\left(v\right)}=\mathbb{E}e^{iv\left(\mu +\gamma U\left(1\right)+\rho T\left(U\left(1\right)\right)+\sigma B_{T\left(U\left(1\right)\right)}\right)}\\
	\indent=e^{iv\mu }{\mathrm{exp} \left\{-\sqrt{-2b_U\left(iv\gamma -\sqrt{-2b_T\left(iv\rho -\frac{1}{2}v^2{\sigma }^2\right)}\right)}\right\}}$.\\	
	
\end{appendix}

\begin{appendix}
	\appendix{\textbf{A.4: $\tau$-compounded L\textrm{\'{e}}vy-stable subordinator Laplace exponent}}	
	\label{appendix:iv}
	
	\noindent   Let $b_n=B,n\in \mathcal{N}=\left\{1,2,\dots .\right\}$. Then for every $\tau>0$ define $V^{\left(\tau \right)}\left(t\right),t\ge 0$ as the $\tau$-compounded L\textrm{\'{e}}vy-stable subordinator with Laplace exponent denoted by ${\mathrm{\Phi }}_{V^{\left(\tau \right)}}\left(s\right),\ s>0.$  Therefore, for $\tau =n$, the Laplace exponent of $V^{\left(n\right)}(s)$ is given by
	\[{\mathrm{\Phi }}_{V^{\left(n\right)}}\left(s\right)=s^{2^{-n}}\prod^n_{k=1}{{\left(2b_k\right)}^{2^{-k}}}=s^{2^{-n}}\prod^n_{k=1}{{\left(2B\right)}^{2^{-k}}}=s^{2^{-n}}{\left(2B\right)}^{\sum^n_{k=1}{2^{-k}}}=s^{2^{-n}}{\left(2B\right)}^{1-2^{-n}}.\]
	
	\noindent Hence, for $ n\in \mathcal{N}=\left\lbrace 1,2,\dots \right\rbrace  $  we find that
	
	\[ {\mathrm{\Phi }}_{V^{\left(n\right)}}\left(s\right)-{\mathrm{\Phi }}_{V^{\left(n-1\right)}}\left(s\right)=s^{2^{-n}}exp\left(\sum^{n-1}_{k=1}{{\mathrm{ln} \left({\left(2b_k\right)}^{2^{-k}}\right)\ }}\right)\left({\left(2b_n\right)}^{2^{-n}}-s^{2^{-n-1}-2^{-n}}\right).\] 
	
	\indent Thus, for every $\tau>0$, we have\\
	\noindent${\mathrm{\Phi }}_{V^{\left(\tau +dt\right)}}\left(s\right)-{\mathrm{\Phi }}_{V^{\left(\tau \right)}}\left(s\right)=s^{2^{-\tau }}{\mathrm{exp} \left(\int^{\tau -dt}_0{{ln \left({\left(2b_y\right)}^{2^{-y}}\right)\ }dy}\right)\ }\left({\left(2b_{\tau }\right)}^{2^{-\tau }-2^{-\tau -dt}}-0\right)dt\\=s^{2^{-\tau }}{\mathrm{exp} \left(\int^{\tau }_0{{ln \left({\left(2b_y\right)}^{2^{-y}}\right)\ }dy}\right)\ }dt$.\\ 
	\noindent Since 
	\[{\mathrm{\Phi }}_{V^{\left(\tau \right)}}\left(s\right)=\frac{{\mathrm{\Phi }}_{V^{\left(n\right)}}\left(s\right)}{{\mathrm{\Phi }}_{V^{\left(n-1\right)}}\left(s\right)}=s^{2^{-n}-2^{-n-1}}2b_n,\]
	a simple calculation shows that
	\[\frac{{\mathrm{\Phi }}_{V^{\left(n\right)}}\left(s\right)-{\mathrm{\Phi }}_{V^{\left(n-1\right)}}\left(s\right)}{{\mathrm{\Phi }}_{V^{\left(n-1\right)}}\left(s\right)}=s^{2^{-n}(1-2)}2b_n-1.\]  
	
	\noindent Thus, we have $\frac{{\mathrm{\Phi }}_{V^{\left(\tau +dt\right)}}\left(s\right)-{\mathrm{\Phi }}_{V^{\left(\tau \right)}}\left(s\right)}{{\mathrm{\Phi }}_{V^{\left(\tau \right)}}\left(s\right)}=\left(s^{2^{-\tau }}2b_{\tau }-1\right)dt$. Therefore, we find \\$\frac{\partial }{\partial \tau }ln{\mathrm{\Phi }}_{V^{\left(\tau \right)}}\left(s\right)=\left(s^{2^{-\tau }}2b_{\tau }-1\right)\,$ and,   $\,ln{\mathrm{\Phi }}_{V^{\left(\tau \right)}}\left(s\right)-ln{\mathrm{\Phi }}_{V^{\left(0\right)}}\left(s\right)=\int^{\tau }_0{s^{-2^{-y}}2b_ydy}$.

	\noindent  Setting   $V^{\left(0\right)}\left(t\right)=t$,  
	\noindent  then $\mathrm{\ }{\mathrm{\Phi }}_{V^{\left(0\right)}}\left(s\right)=-ln\mathbb{E}e^{-sV^{\left(0\right)}\left(1\right)}=-ln\mathbb{E}e^{-s}=s$. 
	
	Finally, we find\\
	$ln{\mathrm{\Phi }}_{V^{\left(\tau \right)}}\left(s\right)=
	lns+{\mathrm{ln} \left(exp\int^{\tau }_0{\left(s^{-2^{-y}}2b_y-1\right)dy}\right)\ }=ln{\mathrm{\Phi }}_{V^{\left(\tau \right)}}\left(s\right)=\\
	{\mathrm{ln} \left(se^{-\tau +\int^{\tau }_0{s^{-2^{-y}}2b_ydy}}\right)}$,\\
	Or,
	\[{\mathrm{\Phi }}_{V^{\left(\tau \right)}}\left(s\right)=s^{2^{-\tau }} {\left(2B\right)}^{1-2^{-\tau }}, s>0.\]
	
\end{appendix}	

\begin{appendix}
	\appendix{\textbf{A.5: Ch.f. of normal-compound(n)-stable log price process}}	
	\label{appendix:v}
	
	\noindent  Consider a log-price process $L^{\left(n\right)}_t=lnS_t$, $t\ge0,n=2,3,..$  of the form
	\begin{equation}
	\label{log_price_n}
	L^{\left(n\right)}_t=L^{\left(n\right)}_0+\mu t+\sum^n_{k=1}{{\gamma }_kV^{\left(k\right)}\left(t\right)}+\sigma B_{T(V^{\left(n\right)}\left(t\right))},t\ge 0,
	\end{equation}
	where the 
	$V^k\left(1\right)\sim \text{L\textrm{\'{e}}vy-compound-stable}$, and  $T\left(1\right)\sim\text{L\textrm{\'{e}}vy-stable}\left(b_T\right)$ 
	are independent processes,
	and $B_s$, $s\ge0$ is a standard Brownian motion.
	\noindent Denote  \[{\mathrm{\Lambda }}^{\left(\mathrm{n}\right)}\mathrm{:}=L^{\left(n\right)}_1-L^{\left(n\right)}_0=\mu +\sum^n_{k=1}{{\gamma }_kV^{\left(n\right)}\left(1\right)}+\sigma B_{T\left(V^{\left(n\right)}\left(1\right)\right)}.\] Then, the Ch.f. of ${\mathrm{\Lambda }}^{\left(\mathrm{n}\right)}$, $n=2,3,\dots$ is given by

	\[{\varphi }_{{\mathrm{\Lambda }}^{\left(\mathrm{n}\right)}}\left(v\right)=\mathbb{E}e^{iv{\mathrm{\Lambda }}^{\left(\mathrm{n}\right)}}=\]\\
	\footnotesize${\mathrm{exp}\left\{iv\mu-\sqrt{-2b_1\left(iv{\gamma }_1-\sqrt{-2b_2\left(\dots \sqrt{-2b_{n-1}\left(iv{\gamma }_{n-1}-\sqrt{-2b_n\left(iv{\gamma }_n-\frac{1}{2}v^2{\sigma }^2\right)}\right)}\dots \right)}\right)}\right\}}.$

	\normalsize
	\noindent For simplicity, we consider when $n=3$, we have:\\
	${\mathrm{\Lambda }}^{\left(\mathrm{3}\right)}\mathrm{:}=L^{\left(3\right)}_1-L^{\left(3\right)}_0=\mu +{\gamma }_1V^{\left(1\right)}\left(1\right)+{\gamma }_2V^{\left(2\right)}\left(1\right)+{\gamma }_3V^{\left(3\right)}\left(1\right)+\sigma B_{T\left(V^{\left(3\right)}\left(1\right)\right)}=\mu +{\gamma }_1U^{\left(1\right)}\left(1\right)+{\gamma }_2U^{\left(2\right)}\left(U^{\left(1\right)}\left(1\right)\right)+{\gamma }_3U^{\left(3\right)}\left(U^{\left(2\right)}\left(U^{\left(1\right)}\left(1\right)\right)\right)+\sigma B_{T\left(U^{\left(3\right)}\left(U^{\left(2\right)}\left(U^{\left(1\right)}\left(1\right)\right)\right)\right).}$ 
	
	Thus the Ch.f. of ${\mathrm{\Lambda }}^{\left(\mathrm{3}\right)}$ is given by\\
	\noindent ${\varphi }_{{\mathrm{\Lambda }}^{\left(\mathrm{3}\right)}}\left(v\right)=\mathbb{E}e^{iv{\mathrm{\Lambda }}^{\left(\mathrm{3}\right)}}=e^{iv\mu }{\mathbb{E}}_{U^{\left(1\right)}\left(1\right)=u}{\left\{\mathbb{E}{\mathrm{exp} \left(iv\left( \begin{array}{c}
			{\gamma }_1+{\gamma }_2U^{\left(2\right)}\left(1\right)+ \\ 
			+{\gamma }_3U^{\left(3\right)}\left(U^{\left(2\right)}\left(1\right)\right)+ \\ 
			+\sigma B_{T\left(U^{\left(3\right)}\left(U^{\left(2\right)}\left(1\right)\right)\right)} \end{array}
			\right)\right)\ }\right\}}^u$  
	
	\noindent From $U^{\left(k\right)}\left(1\right)\sim L\textrm{\'{e}}vystable\left(b_k\right),k=1,2,\dots$, and\\ ${\varphi }_{\mathrm{\Lambda }}\left(v\right)=\mathbb{E}e^{iv\mathrm{\Lambda }}=e^{iv\mu }{\mathrm{exp} \left\{-\sqrt{-2b_U\left(iv\gamma -\sqrt{-2b_T\left(iv\rho -\frac{1}{2}v^2{\sigma }^2\right)}\right)}\right\}\ }$  with\\ $\mathrm{\Lambda }\mathrm{:}=L_1-L_0=\mu +\gamma U\left(1\right)+\rho T\left(U\left(1\right)\right)+\sigma B_{T\left(U\left(1\right)\right)}$, it follows that\\
	$\mathbb{E}{\mathrm{exp} \left(iv\left( \begin{array}{c}
		{\gamma }_1+{\gamma }_2U^{\left(2\right)}\left(1\right)+ \\ 
		+{\gamma }_3U^{\left(3\right)}\left(U^{\left(2\right)}\left(1\right)\right)+ \\ 
		+\sigma B_{T\left(U^{\left(3\right)}\left(U^{\left(2\right)}\left(1\right)\right)\right)} \end{array}
		\right)\right)\ }=e^{iv{\gamma }_1}{\mathrm{exp} \left\{-\sqrt{-2b_2\left(iv{\gamma }_2-\sqrt{-2b_3\left(iv{\gamma }_3-\frac{1}{2}v^2{\sigma }^2\right)}\right)}\right\}\ }.$  
	
	\noindent Thus, ${\varphi }_{{\mathrm{\Lambda }}^{\left(\mathrm{3}\right)}}\left(v\right)=\mathbb{E}e^{iv{\mathrm{\Lambda }}^{\left(\mathrm{3}\right)}}=e^{iv\mu }{\mathbb{E}}_{U^{\left(1\right)}\left(1\right)=u}{\left\{\mathbb{E}{\mathrm{exp} \left(iv\left( \begin{array}{c}
			{\gamma }_1+{\gamma }_2U^{\left(2\right)}\left(1\right)+ \\ 
			+{\gamma }_3U^{\left(3\right)}\left(U^{\left(2\right)}\left(1\right)\right)+ \\ 
			+\sigma B_{T\left(U^{\left(3\right)}\left(U^{\left(2\right)}\left(1\right)\right)\right)} 
			\end{array}
			\right)\right)\ }\right\}}^u\\=e^{iv\mu }\mathbb{E}\left[{\mathrm{exp} \left(\left\{iv{\gamma }_1-\sqrt{-2b_2\left(iv{\gamma }_2-\sqrt{-2b_3\left(iv{\gamma }_3-\frac{1}{2}v^2{\sigma }^2\right)}\right)}\right\}U^{\left(1\right)}\left(1\right)\right)\ }\right]\\=e^{iv\mu }\mathbb{E}\left[{\mathrm{exp} \left(i\left(\frac{iv{\gamma }_1-\sqrt{-2b_2\left(iv{\gamma }_2-\sqrt{-2b_3\left(iv{\gamma }_3-\frac{1}{2}v^2{\sigma }^2\right)}\right)}}{i}\right)U^{\left(1\right)}\left(1\right)\right)\ }\right]$. 
	
	\noindent We know the Ch.f of $U\sim L\textrm{\'{e}}vystable\left(b\right)$ is  ${\varphi }_{U}\left(u\right)=\mathbb{E}e^{iuU}={\mathrm{exp} \left\{-\sqrt{-2ibu}\right\}},u\in R$. This leads to \\
	${\varphi }_{{\mathrm{\Lambda }}^{\left(\mathrm{3}\right)}}\left(v\right)=e^{iv\mu }\left[{\mathrm{exp} \left\{-\sqrt{-2ib_1\frac{iv{\gamma }_1-\sqrt{-2b_2\left(iv{\gamma }_2-\sqrt{-2b_3\left(iv{\gamma }_3-\frac{1}{2}v^2{\sigma }^2\right)}\right)}}{i}}\right\}\ }\right]\\={\mathrm{exp} \left\{iv\mu -\sqrt{-2b_1\left(iv{\gamma }_1-\sqrt{-2b_2\left(iv{\gamma }_2-\sqrt{-2b_3\left(iv{\gamma }_3-\frac{1}{2}v^2{\sigma }^2\right)}\right)}\right)}\right\}}$  
	
	Consequently, for $n>3$ we find\\
	
	\[{\varphi }_{{\mathrm{\Lambda }}^{\left(\mathrm{n}\right)}}\left(v\right)=\mathbb{E}e^{iv{\mathrm{\Lambda }}^{\left(\mathrm{n}\right)}}=\]\\
	\footnotesize \[ {\mathrm{exp}\left\{iv\mu-\sqrt{-2b_1\left(iv{\gamma }_1-\sqrt{-2b_2\left(\dots \sqrt{-2b_{n-1}\left(iv{\gamma }_{n-1}-\sqrt{-2b_n\left(iv{\gamma }_n-\frac{1}{2}v^2{\sigma }^2\right)}\right)}\dots \right)}\right)}\right\}}.\]		
	
\end{appendix}	
\normalsize

\begin{appendix}
	\appendix{\textbf{A.6: Double-gamma subordinator moment-generating function}}	
	\label{appendix:vii}

	\noindent If  $T\left(1\right)\sim Gamma\left({\alpha }_T,{\lambda }_T\right),$  ${\alpha }_T,>0,\ {\lambda }_T\ >0$,  and $U\left(1\right)\sim Gamma\left({\alpha }_U,{\lambda }_U\right)$, then  we have the following representation for the MGF for the double gamma subordinator $T(U(t))$: \\ $M_{T\left(U\left(1\right)\right)}\left(v\right)={\mathbb{E}}_{U\left(1\right)=u}{\left(\mathbb{E}e^{vT\left(1\right)}\right)}^u={\mathbb{E}}_{U\left(1\right)=u}{\left({\left(1-\frac{v}{{\lambda }_T}\right)}^{-{\alpha }_T}\right)}^u\\
	=\frac{{\lambda }^{{\alpha }_U}_U}{\mathrm{\Gamma }({\alpha }_U)}\int^{\infty }_0{e^{\left(-{\alpha }_T{\mathrm{ln} \left(1-\frac{v}{{\lambda }_T}\right)\ }\right)u}u^{{\alpha }_U-1}e^{-{\lambda }_Uu}du}=\frac{{\lambda }^{{\alpha }_U}_U}{{\left({\lambda }_U+{\alpha }_T{\mathrm{ln} \left(1-\frac{v}{{\lambda }_T}\right)\ }\right)}^{{\alpha }_U}},\ 0<v<{\lambda }_T$.  Thus, \[M_{T\left(U\left(t\right)\right)}\left(v\right)={\left(M_{T\left(U\left(1\right)\right)}\left(v\right)\right)}^t={\left(1+\frac{{\alpha }_T}{{\lambda }_U}{\mathrm{ln} \left(1-\frac{v}{{\lambda }_T}\right)\ }\right)}^{-{\alpha }_Ut}.\] \\ 
	Note that we must have  $\left(i\right)\ 1-\frac{v}{{\lambda }_T}>0$ and $\left(ii\right)\ 1+\frac{{\alpha }_T}{{\lambda }_U}{\mathrm{ln} \left(1-\frac{v}{{\lambda }_T}\right)\ }>0$. Therefore, the domain of $M_{T\left(U\left(t\right)\right)}\left(v\right)$ is,  $0<v<{\mathrm{min} \left({\lambda }_T,{\lambda }_T\left(1-{\mathrm{exp} \left(\ -\frac{{\lambda }_U}{{\alpha }_T}\right)\ }\right)\right)\ }={\lambda }_T\left(1-{\mathrm{exp} \left(\ -\frac{{\lambda }_U}{{\alpha }_T}\right)\ }\right)$. \\

\end{appendix}		

\begin{appendix}
	\appendix{\textbf{A.7: Variance gamma-gamma L\'{e}vy\ process density and characteristic function}}	
	\label{appendix:vi}
	
	\noindent Let  $L_t=L_0+\mu t+\gamma U\left(t\right)+\rho V\left(t\right)+\sigma B_{V\left(t\right)}, \,\,t\ge0$ be a variance-gamma-gamma L\'{e}vy process. Then its distribution is determined by the unit increment
	\noindent $\mathrm{\Lambda }=L_1-L_0=\mu +\gamma U\left(1\right)+\rho V()+\sigma B_{V(1)}.$ The pdf of $\mathrm{\Lambda }$ is given by \\
	$f_{\mathrm{\Lambda }}\left(x\right)=\frac{\partial }{\partial x}\mathbb{P}\left(\mathrm{\Lambda }\mathrm{\le }\mathrm{x}\right)=\frac{\partial }{\partial x}\int^{\infty }_0{\mathbb{P}\left(\mu +\gamma u+\rho T\left(u\right)+\sigma \sqrt{T\left(u\right)}N(0,1)\mathrm{\le }\mathrm{x}\right)}f_{U\left(1\right)}\left(u\right)du\\=\frac{\partial }{\partial x}\int^{\infty }_0{\left(\int^{\infty }_0{\mathbb{P}\left(N\left(0,1\right)\mathrm{\le }\frac{\mathrm{x-}\mu -\gamma u-\rho y}{\sigma \sqrt{y}}\right)f_{T(u)}\left(y\right)dy}\right)}f_{U\left(1\right)}\left(u\right)du\\=\int^{\infty }_0{\left(\int^{\infty }_0{f_{N\left(0,1\right)}\left(\frac{\mathrm{x-}\mu -\gamma u-\rho y}{\sigma \sqrt{y}}\right)f_{T(u)}\left(y\right)dy}\right)}f_{U\left(1\right)}\left(u\right)du\\=\int^{\infty }_0{\left(\int^{\infty }_0{\frac{1}{\sqrt{2\pi }}e^{-\ \frac{{\left(\mathrm{x-}\mu -\gamma u-\rho y\right)}^2}{2{\sigma }^2y}}f_{T(u)}\left(y\right)dy}\right)}f_{U\left(1\right)}\left(u\right)du$
	
	\noindent Next, because  $T\left(u\right)\sim Gamma\left({\alpha }_T\,u,{\lambda }_T\right)$ and $U\left(1\right)\sim Gamma\left({\alpha }_U,{\lambda }_U\right)$, it follows that\\
	$f_{\mathrm{\Lambda }}\left(x\right)=\int^{\infty }_0{\left(\int^{\infty }_0{\frac{1}{\sqrt{2\pi }}e^{-\ \frac{{\left(\mathrm{x-}\mu -\gamma u-\rho y\right)}^2}{2{\sigma }^2y}}\frac{{\lambda }^{{\alpha }_Tu}_T}{\mathrm{\Gamma }({\alpha }_{Tu})}y^{{\alpha }_{Tu}-1}e^{-{\lambda }_Ty}dy}\right)}\frac{{\lambda }^{{\alpha }_U}_U}{\mathrm{\Gamma }({\alpha }_U)}u^{{\alpha }_U-1}e^{-{\lambda }_Uu}du\\=\frac{1}{\sqrt{2\pi }}\frac{{\lambda }^{{\alpha }_U}_U}{\mathrm{\Gamma }({\alpha }_U)}\int^{\infty }_0{\left(\int^{\infty }_0{e^{-\ \frac{{\left(\mathrm{x-}\mu -\gamma u-\rho y\right)}^2}{2{\sigma }^2y}}y^{{\alpha }_{Tu}-1}e^{-{\lambda }_Ty}dy}\right)\frac{{\lambda }^{{\alpha }_Tu}_T}{\mathrm{\Gamma }({\alpha }_{Tu})}}u^{{\alpha }_U-1}e^{-{\lambda }_Uu}du.$
	
	The expression for the pdf   $f_{\mathrm{\Lambda }}\left(x\right),x\in R$ is computationally intractable in view of the two integrals in the formula. The Ch.f. of $\mathrm{\Lambda }$, ${\varphi }_{\mathrm{\Lambda }}\left(v\right)=\mathbb{E}e^{iv\mathrm{\Lambda }},v\in R$, has the form \\
	
	\noindent ${\varphi }_{\mathrm{\Lambda }}\left(v\right)={\mathbb{E}}_{U\left(1\right)=u}e^{iv\left(\mu +\gamma u\right)}\mathbb{E}e^{iv\left(\rho T(u)+\sigma B_{T\left(u\right)}\right)}={\mathbb{E}}_{U\left(1\right)=u}e^{iv\left(\mu +\gamma u\right)}{\left(\mathbb{E}e^{iv\left(\rho T\left(1\right)+\sigma B_{T\left(1\right)}\right)}\right)}^u.$  
	
	\noindent Note that \\
	\noindent $\mathbb{E}e^{iv\left(\rho T\left(1\right)+\sigma B_{T\left(1\right)}\right)}={\mathbb{E}}_{T\left(1\right)=y}e^{iv\rho y}e^{-\frac{1}{2}v^2{\sigma }^2y}=\int^{\infty }_0{e^{-\left(-iv\rho +\frac{1}{2}v^2{\sigma }^2\right)y}}\frac{{\lambda }^{{\alpha }_T}_T}{\mathrm{\Gamma }\left({\alpha }_T\right)}y^{{\alpha }_T-1}e^{-{\lambda }_Ty}dy\\=\int^{\infty }_0{\frac{{\lambda }^{{\alpha }_T}_T}{\mathrm{\Gamma }\left({\alpha }_T\right)}y^{{\alpha }_T-1}e^{-\left(-iv\rho +\frac{1}{2}v^2{\sigma }^2+{\lambda }_T\right)y}dy}={\left(1-\ iv\frac{\rho }{{\lambda }_T}+\frac{1}{2}v^2\frac{{\sigma }^2}{{\lambda }_T}\right)}^{-{\alpha }_T}.$\\
	
	Conditional on $U(1)$ in ${\varphi }_{\mathrm{\Lambda }}\left(v\right)$ we have\\
	\noindent ${\varphi }_{\mathrm{\Lambda }}\left(v\right)=e^{iv\mu }{\mathbb{E}}_{U\left(1\right)=u}e^{iv\gamma u}{\left(1-\ iv\frac{\rho }{{\lambda }_T}+\frac{1}{2}v^2\frac{{\sigma }^2}{{\lambda }_T}\right)}^{-u{\alpha }_T}\\=e^{iv\mu }\int^{\infty }_0{e^{iv\gamma u}{\left(1-\ iv\frac{\rho }{{\lambda }_T}+\frac{1}{2}v^2\frac{{\sigma }^2}{{\lambda }_T}\right)}^{-u{\alpha }_T}f_{U\left(1\right)}\left(u\right)du}\\=e^{iv\mu }{\left(1-iv\frac{\gamma }{{\lambda }_U}+\frac{{\alpha }_T}{{\lambda }_U}ln\left(1-\ iv\frac{\rho }{{\lambda }_T}+\frac{1}{2}v^2\frac{{\sigma }^2}{{\lambda }_T}\right)\right)}^{-{\alpha }_U}$, 
	
	\noindent and therefore \[{\varphi }_{\mathrm{\Lambda }}\left(v\right)=\mathbb{E}e^{iv\mathrm{\Lambda }}=e^{iv\mu }{\left(1-iv\frac{\gamma }{{\lambda }_U}+\frac{{\alpha }_T}{{\lambda }_U}ln\left(1-\ iv\frac{\rho }{{\lambda }_T}+\frac{1}{2}v^2\frac{{\sigma }^2}{{\lambda }_T}\right)\right)}^{-{\alpha }_U},v\in R.\]
	
	By setting $u=\frac{v}{i}$ we find the following form for the MGF\\
	\[{M}_{\mathrm{\Lambda }}\left(u\right)={\mathrm{e}\mathrm{xp} \left\{\mu u-{\alpha }_U{\mathrm{ln} \left[1-\frac{\gamma }{{\lambda }_U}u+\frac{{\alpha }_T}{{\lambda }_U}ln\left(1-\frac{\rho }{{\lambda }_T}u-\frac{{\sigma }^2}{2{\lambda }_T}u^2\right)\right]\ }\right\}\ },\] \\
	for $u>0$, such that $1-\frac{\gamma }{{\lambda }_U}u+\frac{{\alpha }_T}{{\lambda }_U}ln\left(1-\frac{\rho }{{\lambda }_T}u-\frac{{\sigma }^2}{2{\lambda }_T}u^2\right)>0$ 
	which will be fulfilled for sufficient small  $u>0.$\\	
\end{appendix}		

\begin{appendix}
	\appendix{\textbf{A.8: Compound-(n) gamma subordinator moment-generating function}}	
	\label{appendix:viii}		
	
	\noindent Let   $U^{\left(i\right)}\left(t\right),t\ge 0,\ i=1,\dots ,n,\ n\in \mathcal{N}=\left\{1,2,\dots \right\}$ be a sequence of independent gamma subordinators with $U^{\left(i\right)}\left(1\right)\sim$ gamma $\left({\alpha }_i,{\lambda }_i\right)$, and define $V^{\left(1\right)}\left(t\right)=U^{\left(1\right)}\left(t\right)$, $V^{\left(i+1\right)}\left(t\right)=V^{\left(i\right)}\left(U^{\left(i+1\right)}\left(t\right)\right)$ for $i=1,2,\dots ,n-1$. We shall use the notation $V^{\left(n\right)}\left(t\right)=U^{\left(1\right)}\circ U^{\left(2\right)}\circ \dots \circ U^{\left(n\right)}\left(t\right),\ t\ge 0$.
	
	For simplicity, we consider when $n=3$. From  A.6 we have\\
	$M_{U^{\left(3\right)}\left(U^{\left(2\right)}\left(U^{\left(1\right)}\left(1\right)\right)\right)}\left(v\right)={\mathbb{E}}_{U^{\left(1\right)}\left(1\right)=u}{\left({\left(1+\frac{{\alpha }_3}{{\lambda }_2}{\mathrm{ln} \left(1-\frac{v}{{\lambda }_3}\right)\ }\right)}^{-{\alpha }_2}\right)}^u\\={\mathbb{E}}_{U^{\left(1\right)}\left(1\right)=u}{\left(1+\frac{{\alpha }_3}{{\lambda }_2}{\mathrm{ln} \left(1-\frac{v}{{\lambda }_3}\right)\ }\right)}^{-{\alpha }_2u}=\frac{{\lambda }^{{\alpha }_1}_1}{\mathrm{\gamma }\left({\alpha }_1\right)}\int^{\infty }_0{u^{{\alpha }_1-1}e^{-\left({\lambda }_1+{\alpha }_2{\mathrm{ln} \left(1+\frac{{\alpha }_3}{{\lambda }_2}{\mathrm{ln} \left(1-\frac{v}{{\lambda }_3}\right)\ }\right)\ }\right)u}du}\\= 
	{\left(1+\frac{{\alpha }_2}{{\lambda }_3}{\mathrm{ln} \left(1+\frac{{\alpha }_1}{{\lambda }_2}{\mathrm{ln} \left(1-\frac{v}{{\lambda }_1}\right)\ }\right)\ }\right)}^{-{\alpha }_3}$\\
	Note that the domain of $M_{U^{\left( 3\right)}\left(t \right)}\left(v \right) $ is, $0<v<{\lambda }_3\left(1-{\mathrm{exp} \left(\frac{{\lambda }_2}{{\alpha }_3}\left({\mathrm{exp} \left(-\frac{{\lambda }_1}{{\alpha }_2}\right)\ }-1\right)\right)\ }\right).$  \\
	
	\indent Now, let $V^{\left(n\right)}\left(t\right)=U^{\left(1\right)}\circ U^{\left(2\right)}\circ \dots \circ U^{\left(n\right)}\left(t\right)=V^{\left(n-1\right)}\left(U(t)\right),\ t\ge 0$. Therefore, for any $n\in \mathcal{N}$, we find  
	\small \[M_{V^{\left(n\right)}(1)}\left(v\right)={\left(1+\frac{{\alpha }_{n-1}}{{\lambda }_n}{\mathrm{ln} \left(1+\frac{{\alpha }_{n-2}}{{\lambda }_{n-1}}{\mathrm{ln} \dots {\mathrm{ln} \left(1+\frac{{\alpha }_1}{{\lambda }_2}{\mathrm{ln} \left(1-\frac{v}{{\lambda }_1}\right)\ }\right)\ }\ }\right)\ }\right)}^{-{\alpha }_n},0<v<{\tau }_n,\]

	\normalsize
	\noindent where ${\tau }_n={\lambda }_1\left(1-{\mathrm{exp} \left(-\frac{{\lambda }_2}{{\alpha }_1}\left(\dots \left(1-{\mathrm{exp} \left(-\frac{{\lambda }_{n-1}}{{\alpha }_{n-2}}\left(1-{exp \left(-\frac{{\lambda }_n}{{\alpha }_{n-1}}\right)\ }\right)\right)\ }\right)\ ...\right)\right)\ }\right).$
	
	\noindent Then by the MGF we have\\
	${\left(M_{V^{\left(n\right)}\left(1\right)}\left(v\right)\right)}^{-\frac{1}{{\alpha }_n}}=1+\frac{{\alpha }_{n-1}}{{\lambda }_n}{\mathrm{ln} \left(1+\frac{{\alpha }_{n-2}}{{\lambda }_{n-1}}{\mathrm{ln} \dots {\mathrm{ln} \left(1+\frac{{\alpha }_1}{{\lambda }_2}{\mathrm{ln} \left(1-\frac{v}{{\lambda }_1}\right)\ }\right)\ }\ }\right)}\\=1+\frac{{\alpha }_{n-1}}{{\lambda }_n}\left(-\frac{1}{{\alpha }_{n-1}}\right)lnM_{V^{\left(n-1\right)}\left(1\right)}\left(v\right)=1-\frac{1}{{\lambda }_n}lnM_{V^{\left(n-1\right)}\left(1\right)}\left(v\right),$  \\
	Or, 
	\[M_{V^{\left(n\right)}\left(1\right)}\left(v\right)={\left(1-\frac{1}{{\lambda }_n}lnM_{V^{\left(n-1\right)}\left(1\right)}\left(v\right)\right)}^{-{\alpha }_n}.\]
	
\end{appendix}		

\begin{appendix}
	\appendix{\textbf{A.9: Ch.f of normal-compound variance gamma}}	
	\label{appendix:ix}	
	
	\noindent Let $U^{\left(i\right)}\left(t\right),t\ge 0,\ i=1,\dots ,n,\ n\in \mathcal{N}=\left\{1,2,\dots \right\}$ be a sequence of independent gamma subordinators with $U^{\left(i\right)}\left(1\right)\sim$ Gamma $\left({\alpha }_i,{\lambda }_i\right)$, and define $V^{\left(1\right)}\left(t\right)=U^{\left(1\right)}\left(t\right)$,  $V^{\left(i+1\right)}\left(t\right)=V^{\left(i\right)}\left(U^{\left(i+1\right)}\left(t\right)\right)$ for  $i=1,2,\dots ,n-1.$ Consider next a log-price process $L^{\left(n\right)}_t=lnS_t,\ t\ge 0,n=2,3,..$  of the form
	\[ L^{\left(n\right)}_t=L^{\left(n\right)}_0+\mu t+\sum^n_{k=1}{{\gamma }_k{\tilde{V}}^{\left(k\right)}\left(t\right)}+\sigma B_{V^{\left(n\right)}\left(t\right)},t\ge 0,\] 
	where $\mu \in R,{\gamma }_k\in R,\ k=1,2,..,\sigma >0,{\tilde{V}}^{\left(k\right)}\left(t\right)=U^{\left(k\right)}\left(U^{\left(k-1\right)}\left(\dots \left(U^{\left(1\right)}\left(t\right)\right)\dots \right)\right)$, $k=1,\dots ,n,n\in \mathcal{N},$ and $U^{\left(i\right)}\left(t\right),t\ge 0,\ i=1,\dots,n$ is a sequence of independent gamma subordinators with $U^{\left(i\right)}\left(1\right)\sim Gamma\left({\alpha }_i,{\lambda }_i\right)$.  Denote \textit{ }${\mathit{\Lambda}}^{\left(n\right)}:=L^{\left(n\right)}_1-L^{\left(n\right)}_0=\mu +\sum^n_{k=1}{{\gamma }_kV^{\left(k\right)}\left(1\right)}+\sigma B_{T\left(V^{\left(n\right)}\left(1\right)\right)}$.
	Then, we find Ch.f. of ${\mathrm{\Lambda }}^{\left(\mathrm{n}\right)},n=2,3,\dots.$ 
	
	For, $n=3$ we have\\
	${\varphi }_{{\mathrm{\Lambda }}^{\left(\mathrm{3}\right)}}\left(v\right)=\mathbb{E}e^{iv{\mathrm{\Lambda }}^{\left(\mathrm{3}\right)}}=\mathbb{E}e^{iv\left(\mu +\sum^3_{k=1}{{\gamma }_kV^{\left(k\right)}\left(1\right)}+\sigma B_{V^{\left(3\right)}(1)}\right)}\\
	=\mathbb{E}{\mathrm{exp} \left(iv\left(\mu +\sum^3_{k=1}{{\gamma }_kV^{\left(k\right)}\left(1\right)}+\sigma B_{V^{\left(3\right)}\left(1\right)}\right)\right)\ }\\=e^{iv\mu }{\mathbb{E}}_{U\left(1\right)=u\ }\mathbb{E}{\mathrm{exp} \left(iv\left({\gamma }_1u+{\gamma }_2U^{\left(2\right)}\left(u\right)+{\gamma }_3U^{\left(3\right)}\left(U^{\left(2\right)}\left(u\right)\right)+\sigma B_{U^{\left(3\right)}\left(U^{\left(2\right)}\left(u\right)\right)}\right)\right)\ }\\=e^{iv\mu }{\mathbb{E}}_{U\left(1\right)=u\ }{\left\{\mathbb{E}{\mathrm{exp} \left(iv\left({\gamma }_1+{\gamma }_2U^{\left(2\right)}\left(1\right)+{\gamma }_3U^{\left(3\right)}\left(U^{\left(2\right)}\left(1\right)\right)+\sigma B_{U^{\left(3\right)}\left(U^{\left(2\right)}\left(1\right)\right)}\right)\right)\ }\right\}}^u$.  \\
	
	We know that if $L_t=L_0+\mu t+\gamma U\left(t\right)+\rho V\left(t\right)+\sigma B_{V\left(t\right)},t\ge 0$, then \\  ${\varphi }_{\mathrm{\Lambda }}\left(v\right)=\mathbb{E}e^{iv\mathrm{\Lambda }}=e^{iv\mu }{\left(1-iv\frac{\gamma }{{\lambda }_U}+\frac{{\alpha }_T}{{\lambda }_U}ln\left(1-\ iv\frac{\rho }{{\lambda }_T}+\frac{1}{2}v^2\frac{{\sigma }^2}{{\lambda }_T}\right)\right)}^{-{\alpha }_U},v\in R$.
	
	\noindent Thus, we find\\
	${\varphi}_{{\mathrm{\Lambda}}^{\left(\mathrm{3}\right)}}\left(v\right)
	=e^{iv\mu}{\mathbb{E}}_{U\left(1\right)=u}\mathrm{exp}\mathrm{}\left[iv{\gamma}_1u-{\alpha}_2u{\mathrm{ln} \left(1-iv\frac{{\gamma }_2}{{\lambda}_2}+\frac{{\alpha}_3}{{\lambda}_2}ln\left(1-\ iv\frac{{\gamma }_3}{{\lambda }_3}+\frac{1}{2}v^2\frac{{\sigma }^2}{{\lambda }_3}\right)\right)}\right]$\\
	Then, by having the distribution of $U(1)$, $f_{U\left(1\right)}\left(x\right)=\frac{{\lambda }^{{\alpha }_1}_1}{\mathrm{\Gamma }\left({\alpha }_1\right)}x^{{\alpha }_1-1}e^{-{\lambda }_1x},x\ge 0,$\\ 
	it follows that\\
	${\varphi }_{{\mathrm{\Lambda }}^{\left(\mathrm{3}\right)}}\left(v\right)\\=e^{iv\mu }{\mathbb{E}}_{U\left(1\right)=u\ }\mathrm{exp}\mathrm{}\left[iv{\gamma }_1u-{\alpha }_2u{\mathrm{ln} \left(1-iv\frac{{\gamma }_2}{{\lambda }_2}+\frac{{\alpha }_3}{{\lambda }_2}ln\left(1-\ iv\frac{{\gamma }_3}{{\lambda }_3}+\frac{1}{2}v^2\frac{{\sigma }^2}{{\lambda }_3}\right)\right)\ }\right]\\=e^{iv\mu }{\left[1-iv\frac{{\gamma }_1}{{\lambda }_1}+\frac{{\alpha }_2}{{\lambda }_1}{ln \left(1-iv\frac{{\gamma }_2}{{\lambda }_2}+\frac{{\alpha }_3}{{\lambda }_2}ln\left(1-\ iv\frac{{\gamma }_3}{{\lambda }_3}+\frac{1}{2}v^2\frac{{\sigma }^2}{{\lambda }_3}\right)\right)\ }\right]}^{-{\alpha }_1}$.
	
	Consequently, for any $n\in \mathcal{N}$, we find\\
	\footnotesize
	\[{\varphi }_{{\mathrm{\Lambda }}^{\left(\mathrm{n}\right)}}\left(v\right)=
	e^{iv\mu }{\left(1-iv\frac{{\gamma }_1}{{\lambda }_1}+\frac{{\alpha }_2}{{\lambda }_1}{ln \left(1-\dots -iv\frac{{\gamma }_{n-1}}{{\lambda }_{n-1}}+\frac{{\alpha }_n}{{\lambda }_{n-1}}ln\left(1-\ iv\frac{{\gamma }_n}{{\lambda }_n}+\frac{1}{2}v^2\frac{{\sigma }^2}{{\lambda }_n}\right)\dots \right)\ }\right)}^{-{\alpha }_1}.\]
	
	\normalsize
	
\end{appendix}

\begin{appendix}
	\appendix{\textbf{A.10:  Double-inverse Gaussian subordinator density and moment-generating function.}}	
	\label{appendix:x}

	\noindent Let's consider the case when the subordinators, $T\left(t\right),t\ge 0$ and $U\left(t\right),t\ge 0$ are inverse Gaussian (IG) L\'{e}vy processes, i.e.  $T\left(1\right)\sim IG\left({\lambda }_T,{\mu }_T\right),$ ${\lambda }_T\mathrm{\ >0},{\mu }_T>0,$ and $U\left(1\right)\sim IG\left({\lambda }_U,{\mu }_U\right)$. 
	Note that the MGF of $T\left(u\right),$ 
	
	\noindent 
	\[M_{T\left(u\right)}\left(v\right)={\left(M_{T\left(1\right)}\left(v\right)\right)}^u
	={\mathrm{exp} \left[\frac{\left({\lambda }_Tu^2\right)}{\left({\mu }_Tu\right)}\left(1-\sqrt{1-\frac{2{\left({\mu }_Tu\right)}^2v}{{\lambda }_Tu^2}}\right)\right]\ },\] 
	that is, $T(u)\sim IG\left({\lambda }_Tu^2,{\mu }_Tu\right)$.
	
	Then the pdf $f_{V\left(1\right)}\left(x\right),x>0$  is given by\\
	$f_{V\left(1\right)}\left(x\right)=\frac{\partial }{\partial x}\int^{\infty }_0{\mathbb{P}\left(T\left(u\right)\le x\right)}f_{U\left(1\right)}\left(u\right)du
	=\int^{\infty }_0{f_{T(u)}\left(x\right)}f_{U\left(1\right)}\left(u\right)du\\
	=\int^{\infty }_0{\sqrt{\frac{{\lambda }_Tu^2}{2\pi x^3}}e^{-\frac{{\lambda }_Tu^2{\left(x-{\mu }_Tu\right)}^2}{2{\mu }^2_Tu^2x}}}\sqrt{\frac{{\lambda }_U}{2\pi u^3}}e^{-\frac{{\lambda}_U{\left(u-{\mu}_T\right)}^2}{2{\mu}^2_Uu}}du
	=\int^{\infty}_0{\frac{1}{2\pi}\sqrt{\frac{{\lambda}_T{\lambda }_U}{ux^3}}}{\mathrm{exp} \left(-\frac{{\lambda}_T{\left(x-{\mu}_Tu\right)}^2}{2{\mu }^2_Tx}-\frac{{\lambda}_U{\left(u-{\mu}_T\right)}^2}{2{\mu }^2_Uu}\right)}du.$\\
	Thus, the pdf of $V\left(1\right)$ has the form
	\[f_{V\left(1\right)}\left(x\right)=\frac{1}{2\pi }\sqrt{\frac{{\lambda }_T{\lambda }_U}{x^3}}\int^{\infty }_0{u^{-\frac{1}{2}}{\mathrm{exp} \left(-\frac{{\lambda }_T{\left(x-{\mu }_Tu\right)}^2}{2{\mu }^2_Tx}-\frac{{\lambda }_U{\left(u-{\mu }_T\right)}^2}{2{\mu }^2_Uu}\right)\ }}du,\ x>0.\] 
	
	Next, the MGF of $V\left(1\right)$ is given by\\
	$M_{V(1)}\left(v\right)={\mathbb{E}}_{U\left(1\right)=u}\mathbb{E}e^{vT\left(u\right)}={\mathbb{E}}_{U\left(1\right)=u}{\left(\mathbb{E}e^{vT\left(1\right)}\right)}^u
	={\mathbb{E}}_{U\left(1\right)=u}{\mathrm{exp} \left[\frac{{\lambda }_Tu}{{\mu }_T}\left(1-\sqrt{1-\frac{2{\mu }^2_Tv}{{\lambda }_T}}\right)\right]\ }\\=\mathbb{E}{\mathrm{exp} \left[\frac{{\lambda }_T}{{\mu }_T}\left(1-\sqrt{1-\frac{2{\mu }^2_Tv}{{\lambda }_T}}\right)U(1)\right]\ }
	={\mathrm{exp} \left[\frac{{\lambda }_U}{{\mu }_U}\left(1-\sqrt{1-2\frac{{\mu }^2_U}{{\lambda }_U}\frac{{\lambda }_T}{{\mu }_T}\left(1-\sqrt{1-\frac{2{\mu }^2_Tv}{{\lambda }_T}}\right)}\right)\right]}.$\\	 
	Therefore,
	\[M_{V(1)}\left(v\right)={\mathrm{exp} \left[\frac{{\lambda }_U}{{\mu }_U}\left(1-\sqrt{1-2\frac{{\mu }^2_U}{{\lambda }_U}\frac{{\lambda }_T}{{\mu }_T}\left(1-\sqrt{1-\frac{2{\mu }^2_Tv}{{\lambda }_T}}\right)}\right)\right]}.\] 
	where 
	$v>0,\ 1-\frac{2{\mu }^2_Tv}{{\lambda }_T}>0,1-2\frac{{\mu }^2_U}{{\lambda }_U}\frac{{\lambda }_T}{{\mu }_T}\left(1-\sqrt{1-\frac{2{\mu }^2_Tv}{{\lambda }_T}}\right)>0.$
	That is, 
	\[v<=\frac{{\lambda }_U}{2{\mu }^2_U}\left(\frac{1}{{\mu }_T}-\frac{{\lambda }_U}{4{\mu }^2_U{\lambda }_T}\ \right).\] 
	
	As a result we have
	\[M_{V(1)}\left(v\right)={\mathrm{exp} \left[\frac{{\lambda }_U}{{\mu }_U}\left(1-\sqrt{1-2\frac{{\mu }^2_U}{{\lambda }_U}\frac{{\lambda }_T}{{\mu }_T}\left(1-\sqrt{1-\frac{2{\mu }^2_Tv}{{\lambda }_T}}\right)}\right)\right]}.\] 
	with the restrictions\\
	$0<v<\frac{{\lambda }_T}{2{\mu }^{2}_T}$, if $\frac{{\lambda}_U{\mu }_T}{2{\mu}^2_U{\lambda}_T}\ge 1$,
	and $0<v<{min\left(\frac{{\lambda}_T}{2{\mu}^{2}_T},\frac{{\lambda}_U}{2{\mu}^2_U}\left(\frac{1}{{\mu}_T}-\frac{{\lambda}_U}{4{\mu}^2_U{\lambda }_T}\right)\right)}$, if $\frac{{\lambda}_U{\mu}_T}{2{\mu}^2_U{\lambda}_T}<1$.\\ 
	
\end{appendix}

\begin{appendix}
	\appendix{\textbf{A.11: Normal-compound inverse Gaussian L\'{e}vy\ process density and characteristic function}}	
	\label{appendix:xii}
	
	\noindent Let  $L_t=L_0+\mu t+\gamma U\left(t\right)+\rho V\left(t\right)+\sigma B_{V\left(t\right)}, \,\,t\ge0,$ be a normal-compound inverse Gaussian L\'{e}vy process, then its distribution is determined by the unit increment
	\noindent $\mathrm{\Lambda }=L_1-L_0=\mu +\gamma U\left(1\right)+\rho V()+\sigma B_{V(1)}.$ The pdf of $\mathrm{\Lambda }$ is given by \\
	$f_{\mathrm{\Lambda }}\left(x\right)=\frac{\partial }{\partial x}\mathbb{P}\left(\mathrm{\Lambda }\mathrm{\le }\mathrm{x}\right)=\frac{\partial }{\partial x}\int^{\infty }_0{\mathbb{P}\left(\mu +\gamma u+\rho T\left(u\right)+\sigma \sqrt{T\left(u\right)}N(0,1)\mathrm{\le }\mathrm{x}\right)}f_{U\left(1\right)}\left(u\right)du\\=\frac{\partial }{\partial x}\int^{\infty }_0{\left(\int^{\infty }_0{\mathbb{P}\left(N\left(0,1\right)\mathrm{\le }\frac{\mathrm{x-}\mu -\gamma u-\rho y}{\sigma \sqrt{y}}\right)f_{T(u)}\left(y\right)dy}\right)}f_{U\left(1\right)}\left(u\right)du\\=\int^{\infty }_0{\left(\int^{\infty }_0{f_{N\left(0,1\right)}\left(\frac{\mathrm{x-}\mu -\gamma u-\rho y}{\sigma \sqrt{y}}\right)f_{T(u)}\left(y\right)dy}\right)}f_{U\left(1\right)}\left(u\right)du\\=\int^{\infty }_0{\left(\int^{\infty }_0{\frac{1}{\sqrt{2\pi }}e^{-\ \frac{{\left(\mathrm{x-}\mu -\gamma u-\rho y\right)}^2}{2{\sigma }^2y}}f_{T(u)}\left(y\right)dy}\right)}f_{U\left(1\right)}\left(u\right)du$
	
	Next, because  $T\left(u\right)\sim IG\left({\mu }_Tu,{\lambda }_Tu\right)$ and $U\left(1\right)\sim Gamma\left({\alpha }_U,{\lambda }_U\right)$, we find\\
	$f_{\mathrm{\Lambda }}\left(x\right)=\frac{1}{4\pi^2}\sqrt{\lambda_T\lambda_U}\int_{0}^{\infty}\int_{0}^{\infty}\frac{1}{4t^{\frac{3}{2}}}
	\exp \left(-\frac{x-\mu-\gamma u-\rho t}{2\sigma\sqrt{t}}-\frac{\lambda_T\left(t-u\mu_T \right)^2}{2ut\mu_T^2 }-\frac{\lambda_U\left(u-\mu_U \right) }{2u\mu_U^2}      \right).$\\  
	The expression for the pdf   $f_{\mathrm{\Lambda }}\left(x\right),x\in R$, is computationally intractable in view of the two integrals in the formula. The Ch.f. of $\mathrm{\Lambda }$, ${\varphi }_{\mathrm{\Lambda }}\left(v\right)=\mathbb{E}e^{iv\mathrm{\Lambda }},v\in R$ has the form \\
	
	\noindent ${\varphi }_{\mathrm{\Lambda }}\left(v\right)={\mathbb{E}}_{U\left(1\right)=u}e^{iv\left(\mu +\gamma u\right)}\mathbb{E}e^{iv\left(\rho T(u)+\sigma B_{T\left(u\right)}\right)}={\mathbb{E}}_{U\left(1\right)=u}e^{iv\left(\mu +\gamma u\right)}{\left(\mathbb{E}e^{iv\left(\rho T\left(1\right)+\sigma B_{T\left(1\right)}\right)}\right)}^u.$  
	
	\noindent Note that \\
	\noindent $\mathbb{E}e^{iv\left(\rho T\left(1\right)+\sigma B_{T\left(1\right)}\right)}
	={\mathbb{E}}_{T\left(1\right)=y}e^{iv\rho y}e^{-\frac{1}{2}v^2{\sigma }^2y}
	=\int^{\infty }_0{e^{-\left(-iv\rho +\frac{1}{2}v^2{\sigma }^2\right)y}}
	\sqrt{\frac{{\lambda }_T}{2\pi y^3}}{\mathrm{exp} \left(-\frac{{\lambda }_T{\left(y-{\mu }_T\right)}^2}{2{\mu }^2_Ty}\right)\ }
	dy\\
	=\exp\left( \frac{\lambda_T}{\mu_T}\left( 1-\sqrt{1+\frac{2\mu_T^{2}}{\lambda_T}\left(iv\rho-\frac{1}{2}v^2\sigma^2 \right)    }\right)       \right) .$\\
	
	Conditional on $U(1)$ in ${\varphi }_{\mathrm{\Lambda }}\left(v\right)$ we have\\
	\noindent ${\varphi }_{\mathrm{\Lambda }}\left(v\right)=e^{iv\mu }{\mathbb{E}}_{U\left(1\right)=u}e^{ivu\gamma u}e^{ \frac{\lambda_T}{\mu_T}\left( 1-\sqrt{1+\frac{2\mu_T^{2}}{\lambda_T}\left(iv\rho-\frac{1}{2}v^2\sigma^2 \right)    }\right)       }\\=e^{iv\mu }\int^{\infty }_0{e^{iv\gamma u}e^{ \frac{u\lambda_T}{\mu_T}\left( 1-\sqrt{1+\frac{2\mu_T^{2}}{\lambda_T}\left(iv\rho-\frac{1}{2}v^2\sigma^2 \right)    }\right)       }f_{U\left(1\right)}\left(u\right)du}.\\$ 
	
	\noindent And, therefore, \[{\varphi }_{\mathrm{\Lambda }}\left(v\right)=\mathbb{E}e^{iv\mathrm{\Lambda }}=
	e^{ \,\,iv\mu+\frac{\lambda_U}{\mu_U} \left[ 1-\sqrt{1-\frac{2\mu_U^2}{\lambda_U}\left(\frac{\lambda_T}{\mu_T}\left(  1-\sqrt{1-\frac{2\mu_T^2}{\lambda_T}\left( iv\rho-\frac{1}{2}v^2\sigma^2\right) }\right) +iv\gamma\right) 
		}        
		\right] 
	},v\in R.\]
	
	\noindent The proof of the MGF, ${M}_{\mathrm{\Lambda }}\left(u\right)$, follows by  setting $u=\frac{v}{i}$, and thus is omitted.
	
\end{appendix}

\end{document}